\documentclass[12pt]{article}

\pdfoutput=1
 
\usepackage[top=80pt,bottom=85pt,left=85pt,right=85pt]{geometry}
\usepackage{amssymb}
\usepackage{amsmath}
\usepackage{setspace}
\usepackage{sectsty}
\usepackage{accents}
\usepackage{comment}

\usepackage[usenames]{xcolor}
\usepackage{graphicx,subcaption}
\usepackage{setspace}
\usepackage{float}
\usepackage{booktabs}
\definecolor{link}{rgb}{.8,.15,.1}
\definecolor{pigment}{rgb}{0.36, 0.54, 0.66}
\definecolor{pigment2}{rgb}{0.19, 0.55, 0.91}
\definecolor{pigment3}{rgb}{0.2, 0.2, 0.6}
\definecolor{light-gray}{gray}{0.75}

\usepackage[debug,pageanchor=false,bookmarks=false]{hyperref}
\hypersetup{colorlinks=true,linkcolor=link,citecolor=link,urlcolor=link,linktocpage}

\usepackage{nameref}
\usepackage{zref-xr,zref-user}
\zxrsetup{toltxlabel}

\usepackage{array,multirow,booktabs,longtable}
\usepackage{tikz,mathrsfs}
\usetikzlibrary{arrows,mindmap,backgrounds,decorations.pathmorphing,decorations.pathreplacing,positioning}
\usetikzlibrary{shapes.geometric,shapes.misc,decorations.markings,intersections,decorations.fractals,calc,patterns,through}

\tikzset{
        cvertex/.style={circle,draw=black,inner sep=1pt,outer sep=3pt},
        vertex/.style={circle,fill=black,inner sep=1pt,outer sep=3pt},
        star/.style={circle,fill=yellow,inner sep=0.75pt,outer sep=0.75pt},
        tvertex/.style={inner sep=1pt,font=\scriptsize},
        gap/.style={inner sep=0.5pt,fill=white}}

\tikzstyle{mybox} = [draw=black, fill=blue!10, very thick,
    rectangle, rounded corners, inner sep=10pt, inner ysep=20pt]
\tikzstyle{boxtitle} =[fill=blue!50, text=white,rectangle,rounded corners]

\newcommand{\rr}{\mathbb{R}}
\newcommand{\cc}{\mathbb{C}}
\newcommand{\zz}{\mathbb{Z}}

\DeclareMathOperator{\Tr}{Tr}
\DeclareMathOperator{\SU}{SU}
\DeclareMathOperator{\U}{U}
\DeclareMathOperator{\SO}{SO}

\makeatletter
\newcommand\xleftrightarrow[2][]{%
  \ext@arrow 9999{\longleftrightarrowfill@}{#1}{#2}}
\newcommand\longleftrightarrowfill@{%
  \arrowfill@\leftarrow\relbar\rightarrow}
\makeatother

\makeatletter
\DeclareRobustCommand{\doublewidetilde}[1]{{%
  \mathpalette\double@widetilde{#1}%
}}
\DeclareRobustCommand{\double@widetilde}[2]{%
  \sbox\z@{$\m@th#1\widetilde{#2}$}%
  \ht\z@=.85\ht\z@
  \widetilde{\box\z@}%
}
\makeatother

\makeatletter
\@addtoreset{equation}{section}
\makeatother

\begin{document}

\begin{titlepage}

\begin{center}

\vskip .3in \noindent

{\Large \bf{$\mathcal{N}=1$ superconformal theories with $D_N$ blocks}}

\bigskip

Marco Fazzi and Simone Giacomelli\\

\vskip 1cm
{\small 

Service de Physique Th\'eorique et Math\'ematique and International Solvay Institutes, \\ Universit\'e Libre de Bruxelles, Campus Plaine C.P.~231, B-1050 Bruxelles, Belgium
	
}

\vskip .3cm
{\small \tt \href{mailto:mfazzi@ulb.ac.be}{mfazzi@ulb.ac.be} $\quad$ \href{mailto:simone.giacomelli@ulb.ac.be}{simone.giacomelli@ulb.ac.be}}

\vskip 1cm
     	{\bf Abstract }
\vskip .1in
\end{center}

We study the chiral ring of four-dimensional superconformal field theories obtained by wrapping M5-branes on a complex curve inside a Calabi--Yau three-fold. We propose a field theoretic construction of all the theories found by Bah, Beem, Bobev and Wecht by introducing new building blocks, and prove several $\mathcal{N}=1$ dualities featuring the latter. We match the central charges with those computed from the M5-brane anomaly polynomial, perform the counting of relevant operators and analyze unitarity bound violations. As a byproduct, we compute the exact dimension of ``heavy operators'' obtained by wrapping an M2-brane on the complex curve.\newline

\noindent

\vfill
\eject

\end{titlepage}

\tableofcontents

\section{Introduction} 
\label{sec:intro}

The M5-brane remains one of the most mysterious objects in string theory. In spite of the overwhelming evidence that the worldvolume 
theory of a stack of M5-branes is a six-dimensional SCFT, we still lack a precise definition although several exact results are known. A key strategy to extract information about this theory is to compactify it and study the resulting theory in lower dimension. 
The simplest nontrivial step in this direction is to compactify it on a Riemann surface; this leads to four-dimensional theories. 
As was first found in \cite{gaiotto,agt}, this strategy is particularly fruitful as it allows to improve our knowledge on supersymmetric theories in four dimensions.

The most well-understood setup involves having a stack of $N$ M5-branes wrap a holomorphic curve $\mathcal{C}_g$ (i.e. the Riemann surface) inside a Calabi--Yau manifold. Depending on the details of the background, upon compactification we obtain different four-dimensional field theories with various amounts of supersymmetry: $\mathcal{N}=2$ in the case of a two-fold (the so-called class $\mathcal{S}$ theories of \cite{gaiotto}) and $\mathcal{N}=1$ for a three-fold. In particular, when the three-fold is the direct sum $\mathcal{L}_1 \oplus \mathcal{L}_2 \rightarrow \mathcal{C}_g$ of two line bundles over the Riemann surface which is being wrapped by the branes, we find a vast class of SCFTs labeled by three parameters: the genus $g$ of the Riemann surface, the number $N$ of M5-branes and the first Chern class of the line bundles. These are not all independent, as the Calabi--Yau condition imposes the constraint $p+q=2g-2$, with $p$ and $q$ the degrees of the two line bundles (i.e. $c_1(\mathcal{L}_{1,2})$). We are thus left with one discrete parameter to play with, the so-called \emph{twist parameter} $z=(p-q)/(2g-2) \in \mathbb{Q}$. 

As it turns out, for a given choice of parameters $(N,g,z)$ labeling an SCFT there are several equivalent UV descriptions, which are all dual to each other. In fact one can prove that global symmetries, central charges and superconformal indices only depend on the choice of parameters and the collection of punctures (if any), strongly supporting the conjectured duality \cite{beem-gadde,gadde-maruyoshi-tachikawa-yan,simone,xie,yonekura}. In section \ref{sec:chiral} we will heavily exploit this observation to derive new chiral ring relations for the SCFTs.

In \cite{bbbw-short,bbbw} it was found that all such field theories admit a smooth gravity dual characterized by the very same $z$, generalizing previous work by Maldacena and N\'{u}\~{n}ez \cite{mn1,mn2}, whose solutions correspond to models with $z=0$ ($\mathcal{N}=1$) and $z=\vert 1 \vert$ ($\mathcal{N}=2$) in our notation.\footnote{Some $\mathcal{N}=1$ SCFTs labeled by $z=0$ arise as mass deformations of $\mathcal{N}=2$ theories, as noticed in \cite{sicilian}.} As usual, having the gravity dual at hand translates into the ability of extracting a great deal of information about the SCFT, including the $a$ and $c$ central charges and the spectrum of chiral operators (at least at leading order in $N$). In this paper we will be particularly interested in the ``heavy operator'' one obtains by wrapping an M2-brane on the Riemann surface \cite{bbbw}.

In \cite{bbbw} (building on results of \cite{bah-wecht}) it was also noticed that for $g>1$ and $p,q \geq 0$ there is a clear field theoretic interpretation based on theories of class $\mathcal{S}$: The boundary theory is of so-called generalized quiver type, and is obtained by coupling $2g-2$ copies of the $T_N$ theory via $\mathcal{N}=2$ and $\mathcal{N}=1$ $\SU(N)$ vector multiplets. As in \cite{beem-gadde}, we will call these theories \emph{accessible}. On the contrary, theories labeled by negative $p$ or $q$ (or $g\leq1$) have been deemed \emph{inaccessible}, although there is no argument against their existence. 

In this paper we give a purely four-dimensional field theory interpretation of inaccessible models. The trick is to consider new building blocks besides $T_N$. We follow a proposal given in \cite{agarwal-intriligator-song}, where variants of $T_N$ are obtained by adding adjoint chiral multiplets charged under the global symmetry of $T_N$ and giving them a nilpotent vev, which in turn breaks spontaneously the global symmetry. This procedure leads to three basic models which we will call $D_N$, $\widetilde{D}_N$ and $\doublewidetilde{D}_N$ (recently discussed also in \cite{maruyoshi-song-def,nardoni}). They are obtained by nilpotent Higgsing of, respectively, one, two and three $\SU(N)$ factors of the global symmetry of $T_N$, in the sense just explained. (The nilpotent Higgsing procedure has previously appeared in \cite{gadde-maruyoshi-tachikawa-yan,agarwal-song,agarwal-bah-maruyoshi-song}, and the superpotential deformation it gives rise to was originally studied in \cite{heckman-tachikawa-vafa-wecht}.) From the M-theory perspective, this amounts to compactifying the $(2,0)$ theory (of type $A_{N-1}$) on a thrice-punctured sphere whose punctures (one, two, or three respectively) have been ``closed'' (and cannot be used to glue two blocks via $\mathcal{N}=2$ or $\mathcal{N}=1$ ``tubes'' in a refined pants-decomposition of $\mathcal{C}_g$).

Having a precise field theoretic description at our disposal, we can compute exactly all the quantities that have already been obtained at large $N$ via holography (that is, approximately), for both accessible and inaccessible theories. In particular, we can compute the dimension of chiral operators and the $a$ and $c$ central charges of the latter SCFTs. We find perfect agreement both with the holographic computations of \cite{bbbw} and with the central charges derived from the anomaly polynomial of the parent six-dimensional $(2,0)$ theory. Thus, these computations can be regarded as a test of the proposed field theory interpretation. We also derive several chiral ring relations for the $D_N$, $\widetilde{D}_N$ and $\doublewidetilde{D}_N$ building blocks, and determine the exact dimension of heavy operators related to M2-branes wrapping the Riemann surface in M-theory. 

This paper is organized as follows. In section \ref{sec:charges} we determine the central charges of our field theories and match them with the anomaly polynomial computation of \cite{bbbw}; the main results of the paper are contained in sections \ref{sec:heavy} and \ref{sec:chiral}, in which we respectively discuss heavy operators of our SCFTs and derive the chiral ring relations for $D_N$, $\widetilde{D}_N$ and $\doublewidetilde{D}_N$. In section \ref{sec:bounds} we discuss in detail unitarity bound violations, while in section \ref{sec:index} we perform the counting of relevant operators, finding perfect agreement with the predictions from the superconformal index of these theories studied in \cite{beem-gadde}. In section \ref{sec:conc} we briefly present our conclusions. Finally in appendix \ref{app:dualities} we prove several new $\mathcal{N}=1$ dualities exploited in section \ref{sec:chiral}.

\section{\texorpdfstring{Matching $a$ and $c$ central charges}{Matching a and c central charges}} 
\label{sec:charges}

The trial $a$ and $c$ central charges for accessible and inaccessible four-dimensional $\mathcal{N}=1$ SCFTs have been extracted in \cite{bbbw} by integrating the anomaly polynomial of the $(2,0)$ theory of type $A_{N-1}$ on the Riemann surface. The derivation rests upon the assumption that the infrared R-symmetry (of the four-dimensional theory) is a linear combination of the two $\U(1)$ symmetries naturally rotating the $\cc$ fibers of the two line bundles. One of the coefficients can be set to one while the other, $\epsilon$, has to be determined through $a$-maximization \cite{intriligator-wecht-a}. Denoting the degrees of the line bundles $p=c_1(\mathcal{L}_1)$ and $q=c_1(\mathcal{L}_2)$, the result of this procedure is the following:\footnote{\label{foot:torus} In the case of the torus $g-1$ is zero and formulae \eqref{eq:ac4dN1} should be modified as follows: All occurrences of $(g-1)z$ should be replaced by $(p-q)/2$ whereas all the terms which do not depend on $z$ should be set to zero.}
\begin{subequations}\label{eq:ac4dN1}
\begin{align}
\begin{split} 
a(\epsilon) = &\ \frac{3}{32}(g-1)\left[3(N^3-1)z\epsilon^3-(3N^3-3N)\epsilon^2 - (3N^3-2N-1)z\epsilon \right. + \\ &\ + \left. 3N^3-N-2\right]\ , 
\end{split} \label{eq:apoly} \\
c(\epsilon)= &\ a(\epsilon)-\frac{\Tr R_\epsilon}{16}\ , \label{eq:ccentral}
\end{align}
\end{subequations}
where 
\begin{equation}\label{eq:TrRz}
\Tr R_{\epsilon}=(g-1)(N-1)(1+z\epsilon)
\end{equation}
is the linear R-anomaly of the four-dimensional theory and 
\begin{equation}\label{eq:z}
z=\frac{p-q}{2g-2} \in \mathbb{Q}
\end{equation}
is the so-called \emph{twist parameter}. Notice that exchanging $p$ and $q$ in $z$ just amounts to interchanging the two line bundles (or equivalently to flipping the sign of $\epsilon$). Consequently, without loss of generality we can restrict our attention to the case $p>q$.

In \cite{bbbw} it was found that all theories with $g>1$ and $p,q\geq0$ can be constructed simply by coupling $2g-2$ copies of $T_N$ via $\mathcal{N}=1$ or $\mathcal{N}=2$ vector multiplets. Since $T_N$ has $\mathcal{N}=2$ supersymmetry the geometry is locally of the form $T^*(\mathcal{C}_g)\times \mathbb{C}$, and one of the two line bundles $\mathcal{L}_i$ is identified with the trivial one on $\mathcal{C}_g$, while the other with its cotangent bundle. We can then assign a sign $+$ or $-$ to each $T_N$, depending on whether $\mathcal{L}_1$ or $\mathcal{L}_2$ is identified with the cotangent bundle on the Riemann surface. Gluing together $T_N$ theories of the same sign corresponds physically to an $\mathcal{N}=2$ gauging, whereas gluing $T_N$'s of opposite sign amounts to an $\mathcal{N}=1$ gauging, accompanied by the superpotential term $\Tr\mu_1\mu_2$ (the $\mu_i$ denote the moment maps of the $\SU(N)$ global symmetries of the two $T_N$'s that are gauged together). The degrees $p$ and $q$ then represent the number of $T_N$ blocks with $+$ and $-$ sign, respectively. Clearly this proposal only works when $p,q\geq0$. 

The goal of the present section is to reproduce formulae \eqref{eq:ac4dN1} for inaccessible theories directly in four dimensions, by means of an exact computation. Since it will turn out to be useful in what follows, let us start by evaluating the trial central charges of a $T_N$ block of $+$ sign, assuming the trial R-symmetry is given by the diagonal combination $R_0$ of the $\U(1)$'s rotating the two line bundles plus $\epsilon$ times the antidiagonal combination $F$. In this case we have:
\begin{equation}
R_0 =\frac{1}{2} R_{\mathcal{N}=2} + I_3\ , \qquad F=\frac{1}{2} R_{\mathcal{N}=2} - I_3\ .
\end{equation} 
In the previous formula $R_{\mathcal{N}=2}$ and $I_3$ denote respectively the generator of the $\U(1)_R$ symmetry and the Cartan 
of $\SU(2)_R$. Our convention is that, under the generator $R_0 + \epsilon F$, the moment maps $\mu_i$ of $T_N$ have charge $1-\epsilon$. 

Using the known formulae for $\mathcal{N}=2$ theories \cite{tachikawa-wecht,shapere-tachikawa,kuzenko-theisen},
\begin{equation}
\Tr R_{\mathcal{N}=2}^3=\Tr R_{\mathcal{N}=2}= 48(a-c)\ , \qquad \Tr R_{\mathcal{N}=2}\,I_3^2=2(2a-c)\ ,
\end{equation}
we can easily find the trial $a$ central charge for the $T_N$ theory:
\begin{equation}\label{eq:aTn}
\begin{split}
a_{T_N}(\epsilon) =\ &\frac{3}{128}\left[\epsilon^3(6N^3-18N^2+12)+\epsilon^2(6N-6N^3) \right. + \\ & -\left. \epsilon(6N^3-6N^2-4N+4)+6N^3-12N^2-2N+8\right]\ .
\end{split}
\end{equation} 
The trial $c$ central charge can be obtained by applying \eqref{eq:ccentral}: 
\begin{equation}\label{eq:cTn}
 c_{T_N}(\epsilon)=a_{T_N}(\epsilon)-\frac{\Tr R_{\epsilon}}{16}=a_{T_N}(\epsilon)+\frac{1+\epsilon}{32}(3N^2-N-2)\ .
\end{equation}
One can immediately check that by maximizing \eqref{eq:aTn} and \eqref{eq:cTn} with respect to $\epsilon$ we get the well-known $a$ and $c$ central charges of the $T_N$ theory (see e.g. \cite[Eq. (2.1), Eq. (2.4)]{gaiotto-maldacena} or \cite{chacaltana-distler-1}):
\begin{equation}\label{eq:max-acTn}
a_{T_N}=\frac{N^3}{6}-\frac{5N^2}{16}-\frac{N}{16}+\frac{5}{24}\ , \qquad c_{T_N}=\frac{N^3}{6}-\frac{N^2}{4}-\frac{N}{12}+\frac{1}{6}\ .
\end{equation} 
The trial R-symmetry correctly reduces to the $\U(1)_R$ symmetry of the $\mathcal{N}=1$ subalgebra: 
\begin{equation}
R=\frac{1}{3}R_{\mathcal{N}=2}+\frac{4}{3}I_3\ .
\end{equation}
\newline
Whenever one of the line bundles has negative degree (for $g>0$) -- i.e. in the case of inaccessible theories -- our proposal is that the underlying field theory is a generalized quiver involving $T_N$ theories and another building block (already considered in \cite{agarwal-intriligator-song,maruyoshi-song-def}) which we will call $D_N$ in the present work. The latter is obtained starting from $T_N$, coupling a chiral multiplet $M$ in the adjoint of $\SU(N)$ to the moment map $\mu$ of one of the three $\SU(N)$ flavor symmetries of $T_N$ via $\Tr M \mu$, and finally giving $M$ a nilpotent vev of the form 
\begin{equation}\label{eq:vev}
\langle M \rangle =\begin{bmatrix}
  0 & 1 & 0 & 0 \\
  0 & 0 & \ddots & 0\\
  0 & \dots & 0 & 1\\
  0 & \dots & 0 & 0
\end{bmatrix}\ .
\end{equation}
This theory has a residual $\SU(N) \times \SU(N)$ flavor symmetry inherited from the global symmetry of $T_N$ (the third $\SU(N)$ factor has been completely Higgsed).

By expanding the superpotential term $\Tr M\mu$ around the new vacuum we obtain:
\begin{equation}\label{eq:pote}
\mathcal{W}=\sum_{i=1}^{N-1}\mu_{i+1,i}+\Tr M\mu\ .
\end{equation} 
The first term breaks both $R_{\epsilon}$ and the $\SU(N)$ global symmetry completely, therefore we will have to redefine the trial R-symmetry. All the components of $\mu$ except $N-1$ of them appear in the variation of the above superpotential with respect to $\SU(N)$ (see e.g. \cite{gadde-maruyoshi-tachikawa-yan,maruyoshi-song-def} for the details); consequently, all these chiral multiplets recombine with the $\SU(N)$ currents into long supersymmetry multiplets. The corresponding components of $M$ will then decouple, since the interaction with the $T_N$ fields (dictated by $\Tr M \mu$) has disappeared.  The former are identified with the Goldstone multiplets associated with the spontaneous breaking of $\SU(N)$. Neglecting these decoupled fields, the multiplet $M$ (i.e. its vev plus fluctuations around it) takes the form:
\begin{equation}\label{eq:mvev}
M=\begin{bmatrix}
  0 & 1 & 0 & \dots & 0 \\
  M_{N,2} & 0 & 1 & \dots & 0\\
  M_{N,3} & M_{N,2} & \ddots & \ddots & 0\\
  \vdots & \ddots & \ddots & 0 & 1\\
  M_{N,N} & \dots & M_{N,3} & M_{N,2} & 0
\end{bmatrix}\ .
 \end{equation}
Let us now consider the Cartan generator of $\SU(N)$: 
\begin{equation}\label{eq:gen}
\rho=\begin{bmatrix}
  \frac{N-1}{2} & 0 & \dots & 0 & 0 \\
  0 & \frac{N-3}{2} & 0 & \dots & 0\\
  0 & 0 & \ddots & 0 & 0\\
  0 & \dots & 0 & \frac{3-N}{2} & 0\\
  0 & \dots & & 0 & \frac{1-N}{2}
\end{bmatrix}\ .
\end{equation}
We can immediately see that all the components $\mu_{i+1,i}$ have charge $-1$ and this is the only $\SU(N)$ generator under which 
they have equal nonzero charges. We then combine $R_{\epsilon}$ with $\rho$ to obtain a $\U(1)$ symmetry which is preserved by the nilpotent vev \eqref{eq:vev}, and is thus unbroken along the RG flow.  
Marginality of the superpotential \eqref{eq:pote} requires that the new trial R-symmetry be:\footnote{There is a caveat for $N=2$. 
In fact in this case the terms linear in $\mu$ are actually mass terms, and the correct procedure is to integrate out the massive fields from the superpotential. As it turns out, the massless sector of $T_2$ after nilpotent Higgsing of one $\SU(2)$ flavor group is a chiral multiplet in the bifundamental of $\SU(2)\times \SU(2)$ with charge $-\epsilon$. It is easy to check that the contribution to the central charges of this multiplet plus $M_{2,2}$ reproduces \eqref{eq:newc} with $N=2$.} 
\begin{equation}\label{eq:Repsnew}
R_{\text{new}}=R_{\epsilon}-(1+\epsilon)\rho\ .
\end{equation}
Under this $\U(1)$ group the components of $M$ which do not decouple have charge 
\begin{equation}\label{eq:RnewMcharge}
R_{\text{new}}(M_{N,i})=i(1+\epsilon)\ , \quad i=2,\dots,N\ .
\end{equation}

The trial central charges for the $D_N$ theory can now be computed as follows. The $M_{N,i}$ multiplets give the following contributions:
\begin{subequations}\label{eq:contot}
\begin{equation}\label{eq:con1}
\Tr R_{\text{new}}(M) = \epsilon\left(\frac{N^2}{2}+\frac{N}{2}-1\right)+\frac{N^2}{2}-\frac{N}{2}\ ;
\end{equation}
\begin{multline}\label{eq:con2}
\Tr R^3_{\text{new}}(M) = \epsilon^3\frac{N^2(N+1)^2-4}{4}+\epsilon^2\frac{(N^2-1)(3N^2+2N)}{4} \ + \\ + \epsilon\frac{(N^2-1)(3N^2-2N)}{4}+\frac{N^2(N-1)^2}{4}\ .
\end{multline} 
\end{subequations}
The contribution from the $T_N$ sector is evaluated by including the triangle anomaly $\Tr R_{\mathcal{N}=2}\rho^2$, which is proportional to $\Tr R_{\mathcal{N}=2} \SU(N)^2=-N$. The proportionality coefficient is the embedding index $I_\rho$, which in the present case is equal to 
\begin{equation}\label{eq:embeddingindex}
I_{\rho}=\frac{N(N^2-1)}{6}\ .
\end{equation}
From the above considerations we conclude that:
\begin{equation}
\Tr R_{\text{new}}=\Tr R_{\epsilon}\ ,\qquad \Tr R_{\text{new}}^3=\Tr R_{\epsilon}^3+3(1+\epsilon)^2\Tr R_{\epsilon}\rho^2\ .
\end{equation}
 The terms proportional to $R_{\epsilon}$ reproduce the 't Hooft anomalies \eqref{eq:aTn}, \eqref{eq:cTn} for $T_N$, while the last term gives the contribution 
\begin{equation}\label{eq:con3}
\Tr R_{\epsilon}\rho^2=\frac{I_{\rho}}{2}(1+\epsilon)\Tr R_{\mathcal{N}=2}\SU(N)^2=-\frac{1+\epsilon}{2}\frac{N^2(N^2-1)}{6}\ .\end{equation} 
Notice that, both in formula \eqref{eq:con3} and in the contributions \eqref{eq:contot} from the $M$ multiplet, there are terms proportional to $N^4$ which we do not expect to arise in a theory defined in terms of M5-branes. It is easy, and indeed satisfactory, to check that when we combine together all the terms these unpleasant contributions cancel out.\newline

\begin{figure}[!ht]
\centering
\includegraphics[scale=0.75]{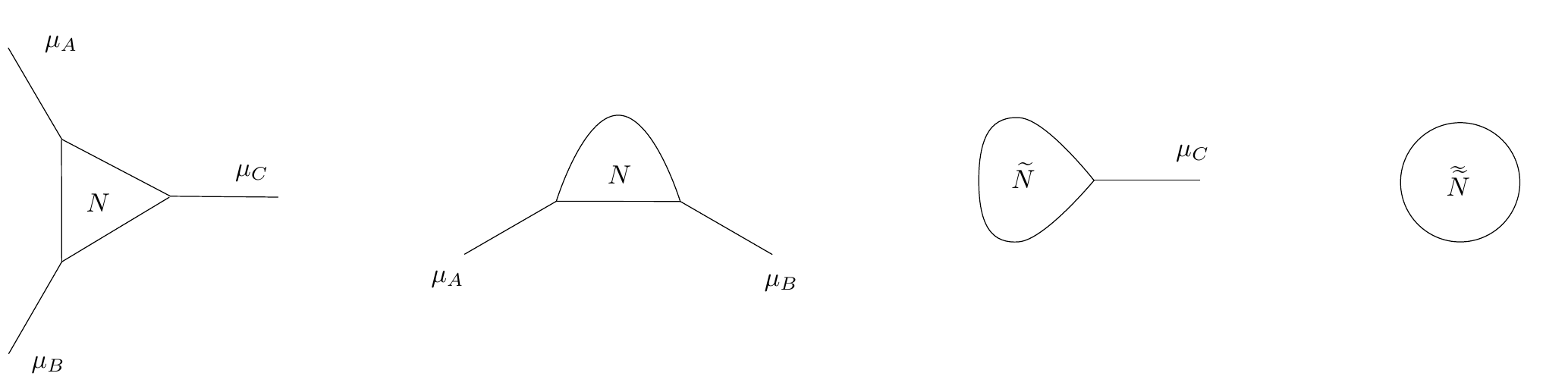}
\caption{The building blocks for $\mathcal{N}=1$ generalized quiver theories. From left to right: $T_N$, $D_N$, $\widetilde{D}_N$, and $\doublewidetilde{D}_N$. The $\mu_X$'s refer to the moment maps associated with the available $\SU(N)_X$ flavor symmetries (represented by legs) in a block. Each leg can be used as a ``tube'' connecting two blocks in a generalized quiver, whereby the flavor symmetries connected by the tube (say $X$ and $Y$) are gauged together in an $\mathcal{N}=2$ or $\mathcal{N}=1$ way, i.e. via the superpotential term $\Tr \Phi (\mu_X -\mu_Y)$ and $\Tr \mu_X \mu_Y$ respectively.}
\label{fig:blocks}
\end{figure}

In order to discuss the case in which the Riemann surface $\mathcal{C}_g$ is a sphere we need to introduce two new building blocks besides $D_N$, which we will call $\widetilde{D}_N$ and $\doublewidetilde{D}_N$ (see also \cite{maruyoshi-song-def} for a related discussion on these theories). These are defined by applying the nilpotent Higgsing procedure to two and three punctures of $T_N$ respectively. We are thus able to construct models with a single $\SU(N)$ global symmetry ($\widetilde{D}_N$) and without nonabelian global symmetries ($\doublewidetilde{D}_N$). See figure \ref{fig:blocks}.

Their central charges can be easily computed by adding to the trial central charges of a $T_N$ theory twice or thrice the contributions \eqref{eq:con1}, \eqref{eq:con2}, and \eqref{eq:con3}.

\subsection{Central charges of all BBBW models}
\label{sub:charges-inac}

We will now propose a precise field theoretic description of all the theories constructed in \cite{bbbw} using the building blocks defined above, and check that the trial central charges reproduce the results obtained from the anomaly polynomial.

\subsubsection{High genus}
\label{subsub:highgenus}

\begin{figure}[!tb]
    \centering
    \begin{minipage}{1\textwidth}
     \captionsetup{type=figure}
        \centering
        \subcaptionbox{$g=3,\ p=4,\ q=0 \Rightarrow z=1$: four $T_N$ blocks of the same sign, chosen to be $+$ (hence the conventional shading, as in \protect\cite{bbbw}). All gaugings are $\mathcal{N}=2$, and the theory actually enjoys $\mathcal{N}=2$ supersymmetry. 
        \label{fig:g3p4}}[.3\columnwidth]{\includegraphics[scale=.2]{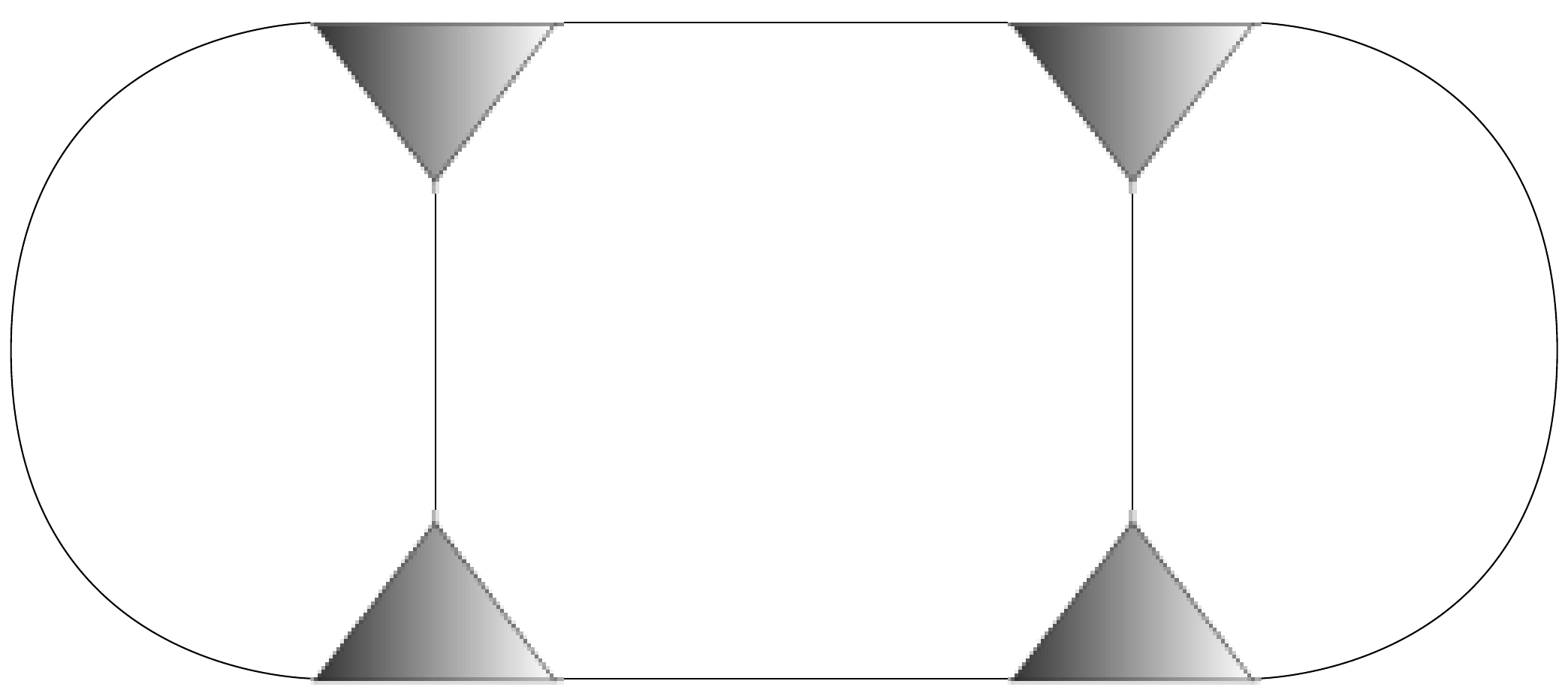}}\hspace{.25cm}
        \subcaptionbox{$g=3,\ p=3,\ q=1 \Rightarrow z=1/2$: three $T_N^+$'s (shaded) and one $T_N^-$ (unshaded). Blocks of different sign are connected by $\mathcal{N}=1$ gaugings. 
        \label{fig:g3p1q3}}[.3\columnwidth]{\includegraphics[scale=.185]{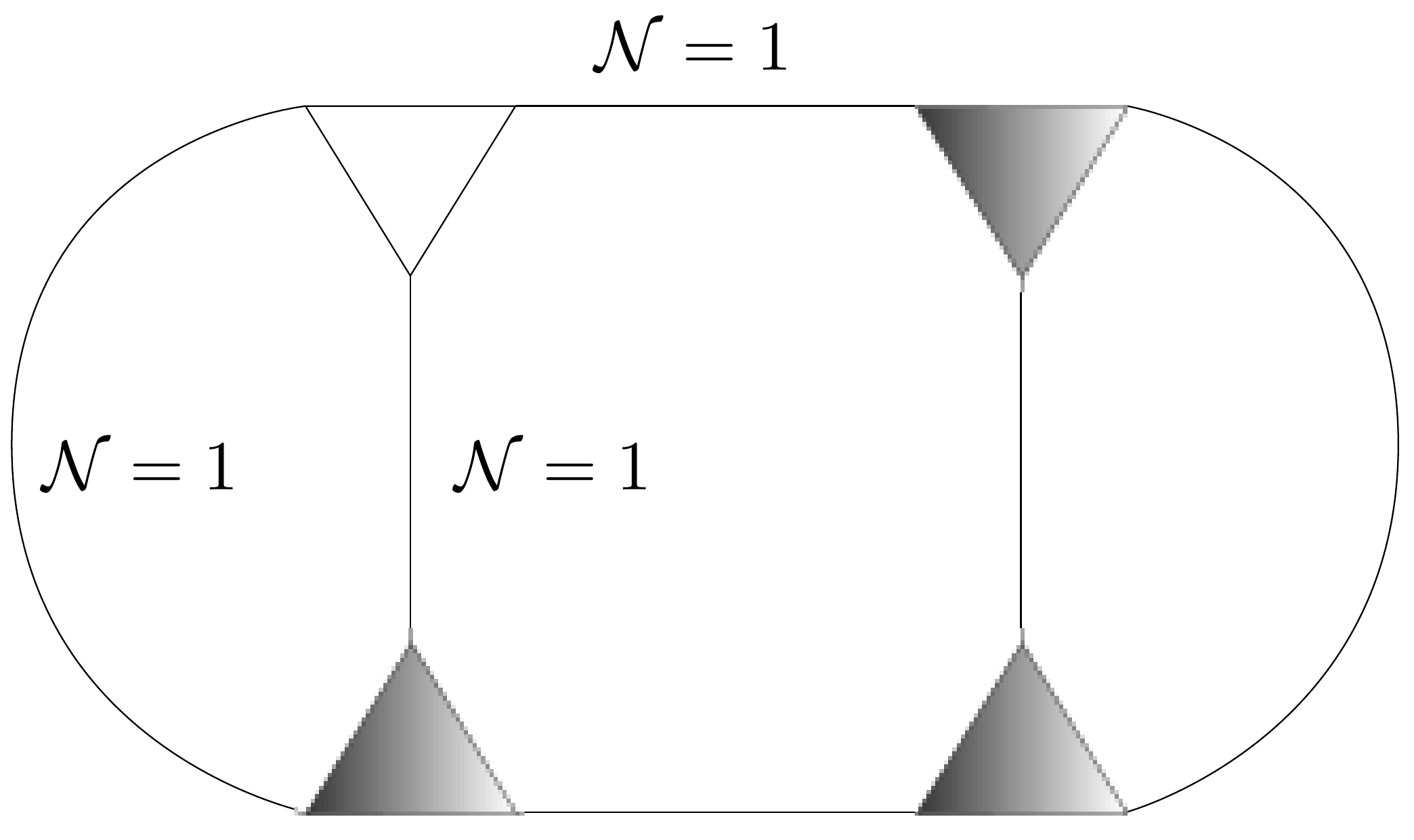}}\hspace{.25cm}
        \subcaptionbox{$g=3,\ p=4+6,\ q=0-6 \Rightarrow z=4$: four $T_N^+$'s and six $D_N$'s, contributing just as many modified tubes. $z$ gets shifted due to the six modified tubes: $z=1 \rightarrow 1+ 6/(3-1) = 4$. All gaugings are $\mathcal{N}=2$, whereas the $D_N$ blocks are intrinsically $\mathcal{N}=1$ theories.
        \label{fig:g36Dn}}[.35\columnwidth]{\includegraphics[scale=.23]{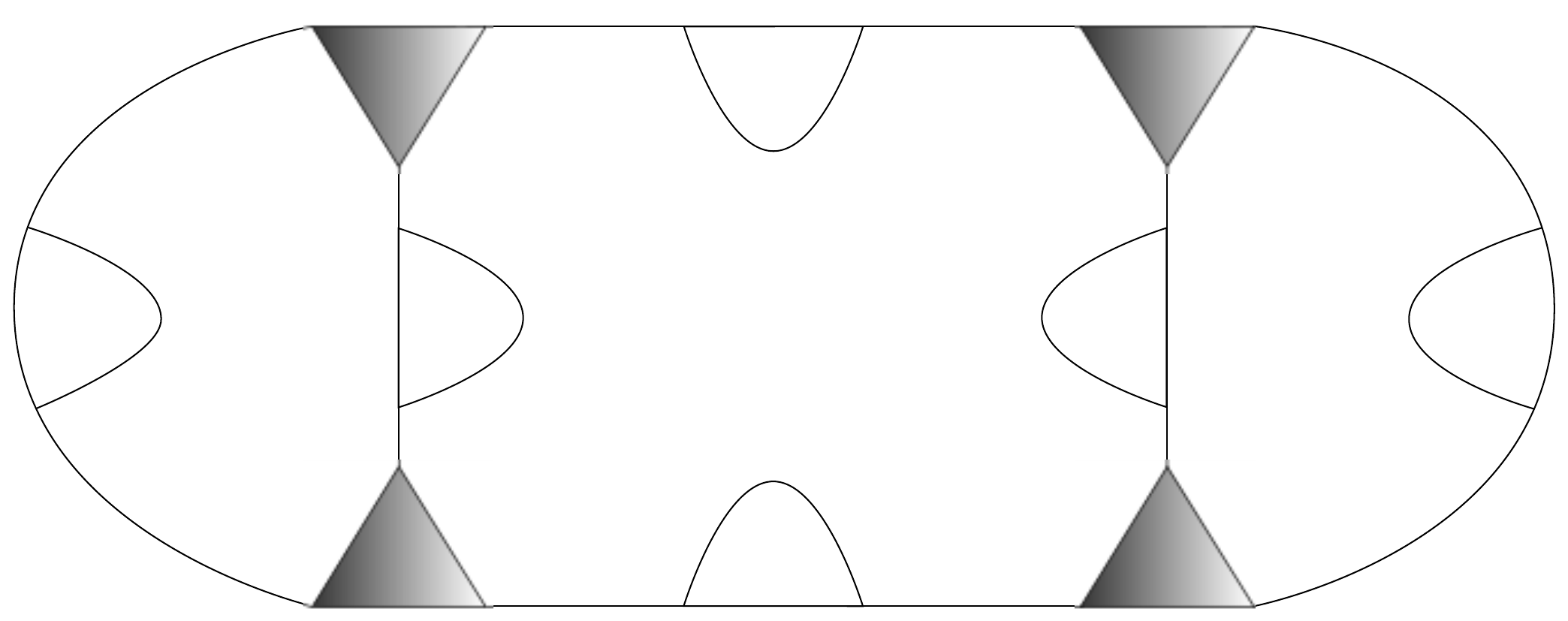}}
    \end{minipage}
    \caption{Examples of $\mathcal{N}=1$ theories engineered by $N$ M5-branes wrapping a holomorphic curve $\mathcal{C}_3$ of genus $g=3$. (There are four other quiver topologies with same genus \protect\cite[Fig. 7]{bbbw}.) The various possibilities depicted in figures \protect\ref{fig:g3p4}, \protect\ref{fig:g3p1q3}, \protect\ref{fig:g36Dn} correspond to different choices of degrees $p$, $q$ of the two line bundles over $\mathcal{C}_3$, such that $2g-2=4=p+q$. To have $p>0, q<0$ we need to introduce some $D_N$ blocks in the generalized quiver.}
\label{fig:g3}
\end{figure}

Let us first consider a generalized quiver of genus $g>1$ obtained by connecting $2g-2$ copies of $T_N$. This is the field theory interpretation 
of the M-theory model with $p=2g-2$ and $q=0$. There are $3g-3$ $\mathcal{N}=2$ $\SU(N)$ vector multiplets whose scalar component has charge $1+\epsilon$ under $R_{\epsilon}$. If we now replace an $\mathcal{N}=2$ vector multiplet with two $\mathcal{N}=2$ vector multiplets coupled to the $D_N$ theory (we call this a \emph{modified tube}), the trial central charges of the theory change by an amount which is equal to the contribution of an $\mathcal{N}=2$ vector multiplet (with scalar component of charge $1+\epsilon$) plus the contribution 
of $D_N$. Combining the formulae given above, it is straightforward to show that the contribution to $a(\epsilon)$ due to the modified tube is 
given by
\begin{equation}\label{eq:newc}
a(\epsilon)_{\text{mod. tube}} = \frac{3}{32}\left[\epsilon^3(3N^3-3)-\epsilon(3N^3-2N-1)\right]\ .
\end{equation}
Adding this to \eqref{eq:apoly} we see that the net effect is simply to replace the parameter $z$ with $z'=z+1/(g-1)$, yielding the trial $a$ central charge of the theory with $p'=p+1$ and $q'=q-1$. Specializing to our case, we now have the trial $a$ central charge of the theory with $p=2g-1$ and $q=-1$. Clearly, this operation can be repeated an arbitrary number $n$ of times, leading to a theory with $p=2g-2+n$ and $q=-n$ for $n \in \mathbb{N}$. Similarly, it is easy to check that the contribution to $\Tr R_{\text{new}}$ of an $\mathcal{N}=2$ vector multiplet plus $D_N$ is $(N-1)\epsilon$, which again corresponds to the shift $z \rightarrow z+1/(g-1)$ in \eqref{eq:ccentral}. This proves that the $a$ and $c$ trial central charges match those computed in \cite{bbbw} for $g>1$ and generic $p,q$. See figure \ref{fig:g3} for a few examples with $g=3$.

\subsubsection{Torus}
\label{subsub:torus}

\begin{figure}[!tb]
    \centering
    \begin{minipage}{1\textwidth}
     \captionsetup{type=figure}
        \centering
        \subcaptionbox{$g=1,\ p=-q=1$: a single $D_N$ block. Its $\SU(N)_B$ and $\SU(N)_C$ flavor symmetries are gauged together. The trial central charge is $a= a_{\mathrm{mod.}\ \mathrm{tube}}$.
        \label{fig:torus-1}}[.45\columnwidth]{\includegraphics[scale=.15]{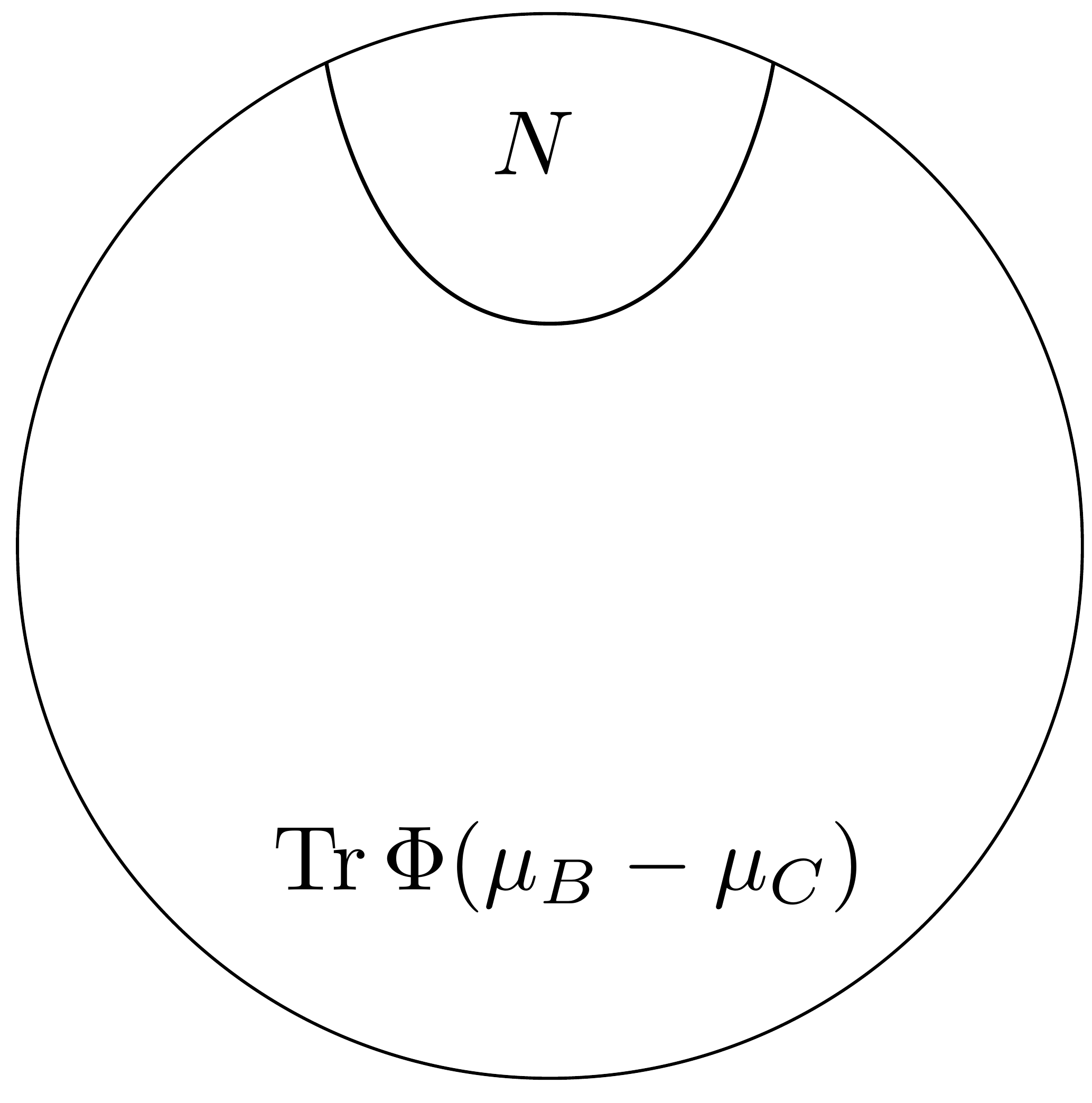}}\hspace{.5cm}
        \subcaptionbox{$g=1,\ p=-q=6$: six $D_N$ blocks. The trial central charge is $a= 6 \, a_{\mathrm{mod.}\ \mathrm{tube}}$.
        \label{fig:torus-6}}[.45\columnwidth]{\includegraphics[scale=.15]{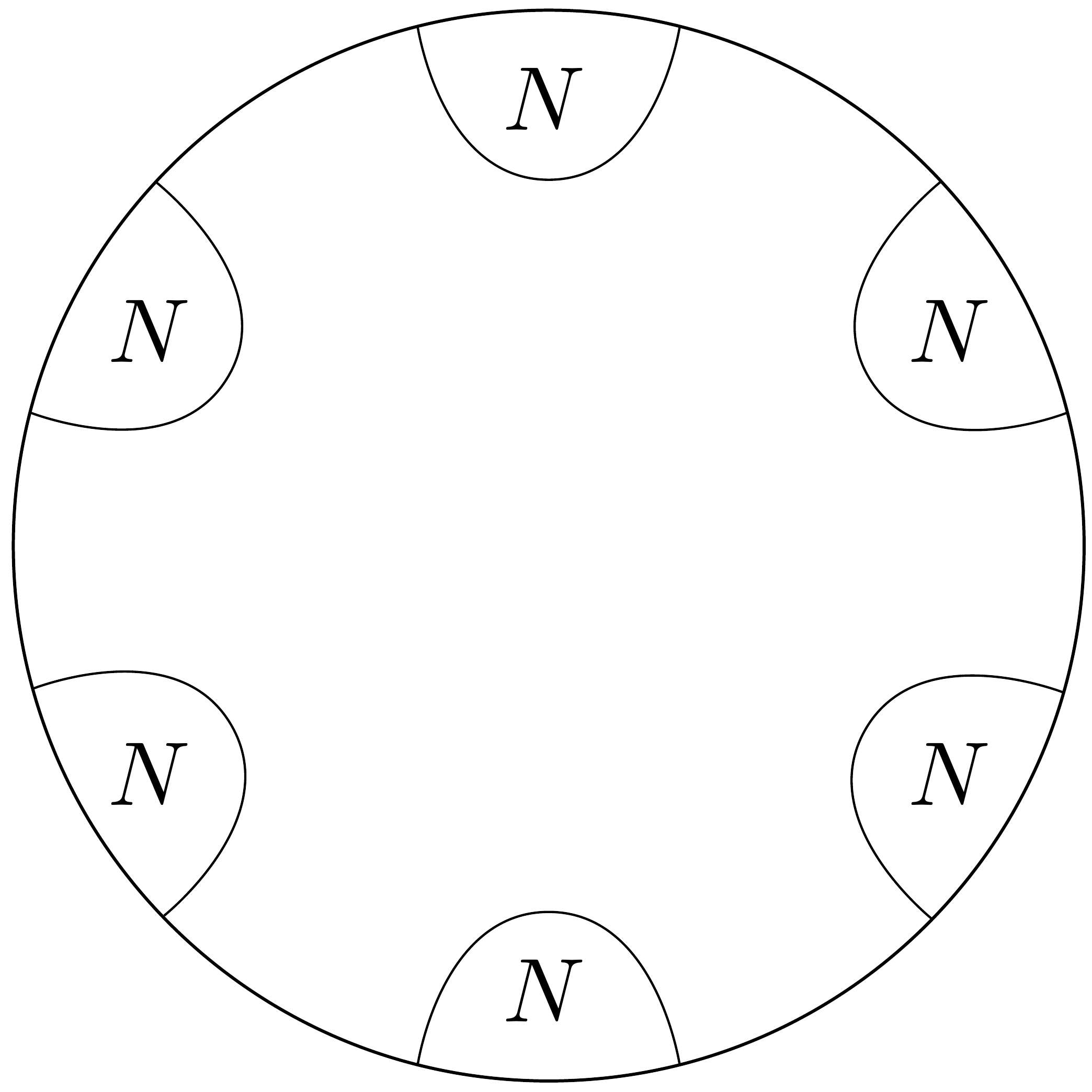}}
    \end{minipage}
    \caption{Examples of $\mathcal{N}=1$ theories engineered by $N$ M5-branes wrapping a torus. Figure \protect\ref{fig:torus-1} shows the ``minimal'' theory consisting of a single $D_N$ block (i.e. $p=1$), its two flavor symmetries being gauged together. Figure \protect\ref{fig:torus-6} shows the $p=6$ case. All gaugings are $\mathcal{N}=2$.}
\label{fig:torus}
\end{figure}

In the $g=1$ case we propose the following field theory interpretation: A circular quiver of $D_N$ theories with $p$ nodes coupled 
through $\mathcal{N}=2$ vector multiplets. Figure \ref{fig:torus} displays two such examples. The trial $a$ central charge is $p$ times the quantity given in \eqref{eq:newc}, which coincides precisely with the result found in \cite{bbbw}, i.e. \eqref{eq:ac4dN1} with $g=1$ (keeping in mind the caveat of footnote \ref{foot:torus}). 

\subsubsection{Sphere}
\label{subsub:sphere}

\begin{figure}[!htb]
    \centering
    \begin{minipage}{1\textwidth}
     \captionsetup{type=figure}
        \centering
        \subcaptionbox{$g=0,\ p=1,\ q=-3 \Rightarrow z=-2$: a single $\doublewidetilde{D}_N$ block. 
        \label{fig:g0-z(-2)}}[.45\columnwidth]{\includegraphics[scale=0.075]{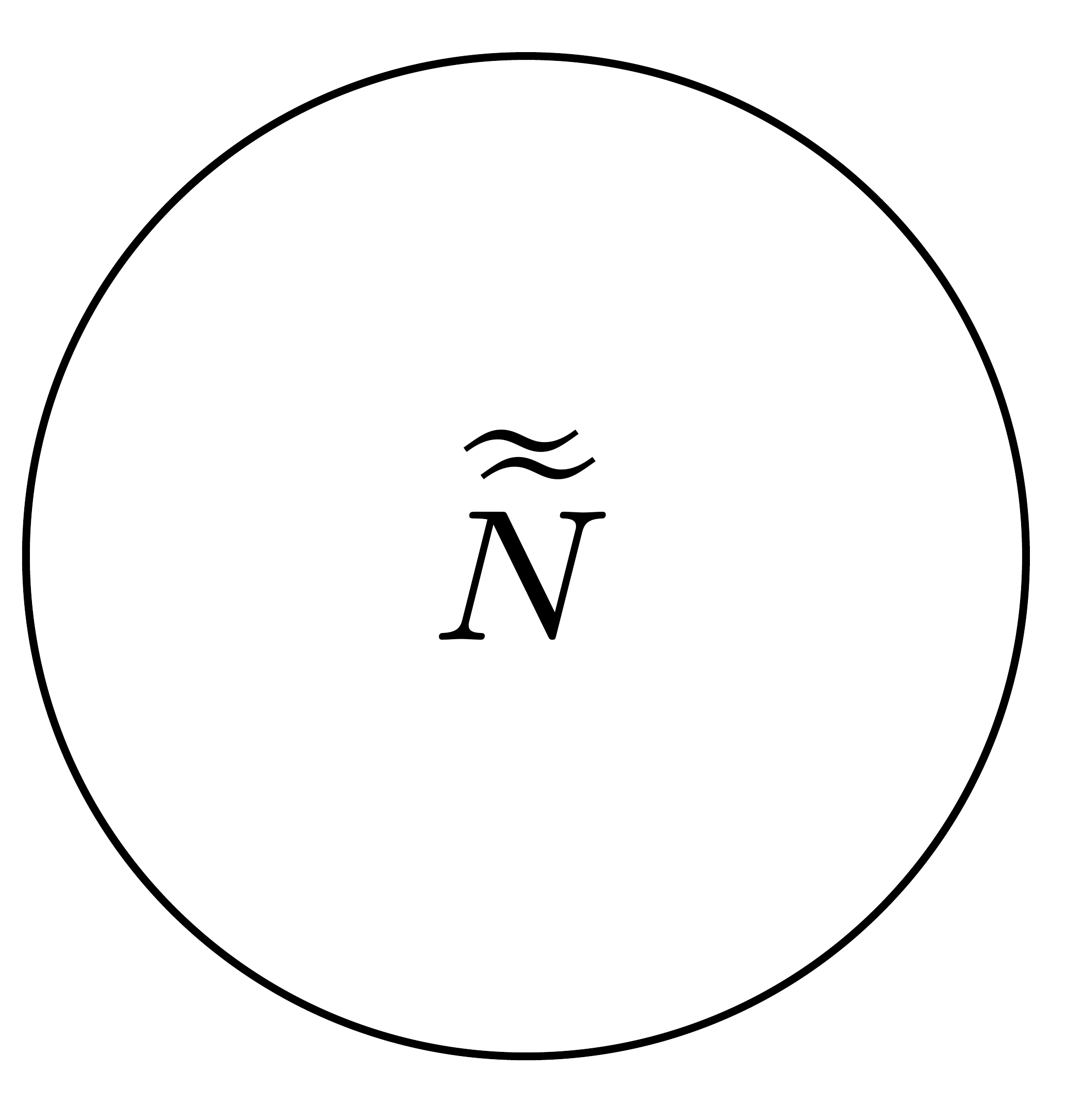}}\hspace{.5cm}
        \subcaptionbox{$g=0,\ p=2,\ q=-4 \Rightarrow z=-3$: two $\widetilde{D}_N$ blocks connected by an $\mathcal{N}=2$ gauging. 
        \label{fig:g0-z(-3)}}[.44\columnwidth]{\includegraphics[scale=0.225]{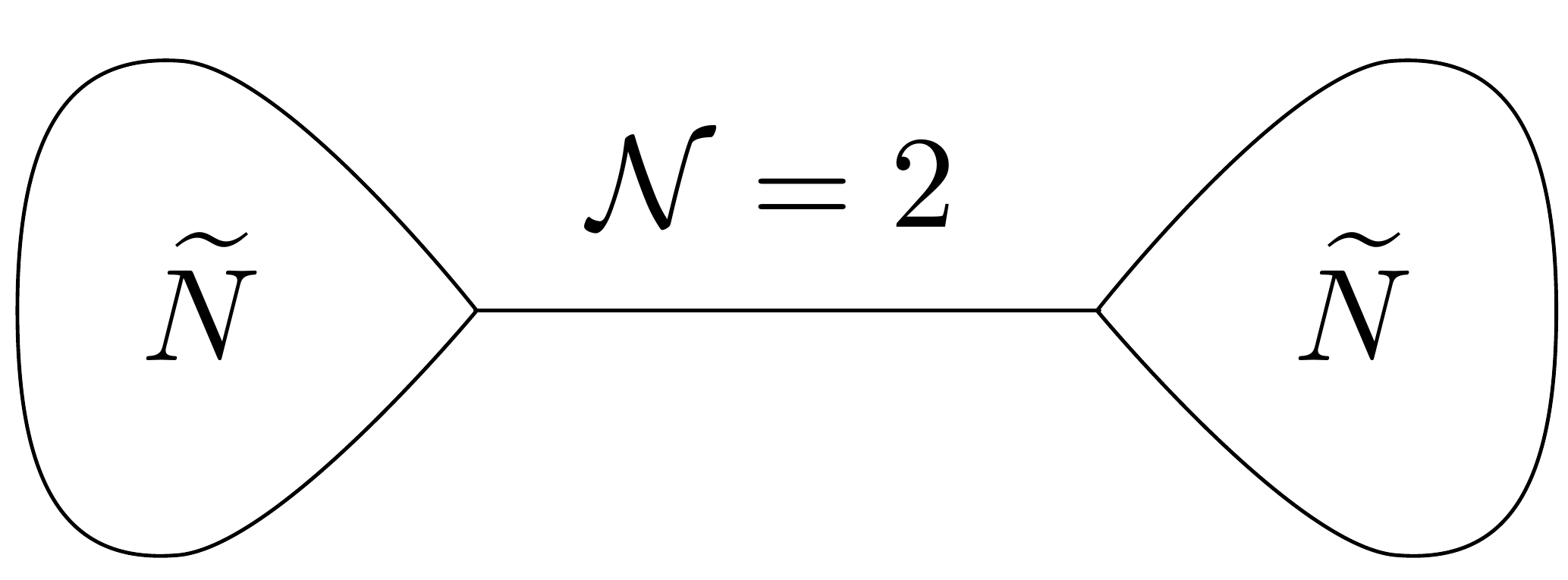}}\vspace{1cm}
        \subcaptionbox{$g=0,\ p=-1+n,\ q=-1-n \Rightarrow z=-n$ for $n\geq 3$: the generic theory on the sphere contains two $\widetilde{D}_N$ blocks at the tails and $-z-3=n-3$ modified tubes between them. $z$ gets shifted due to their contribution: $z=-3 \rightarrow -3+ \frac{n-3}{-1} = -n$. All gaugings are $\mathcal{N}=2$.
        \label{fig:g0-z(-4)and}}[.75\columnwidth]{\includegraphics[scale=0.5]{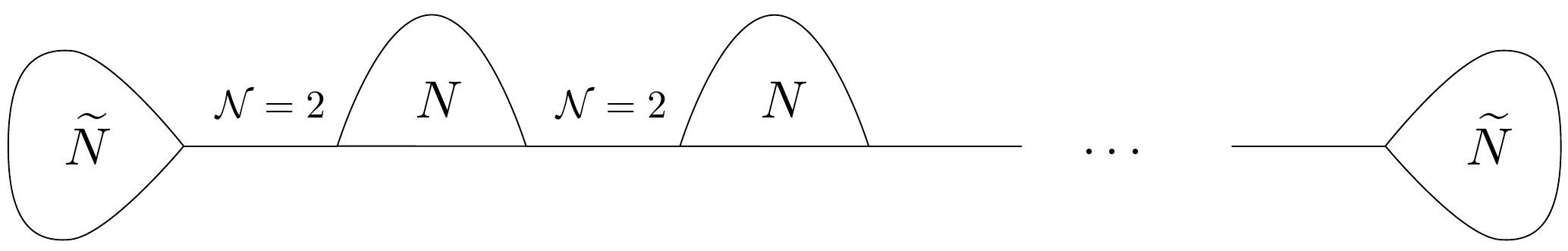}}
    \end{minipage}
    \caption{Examples of $\mathcal{N}=1$ theories engineered by $N$ M5-branes wrapping the sphere. There is no available flavor symmetry (hence no punctures on $\mathcal{C}_0$), as is clear from the absence of external legs in the quivers. The value of $z=\frac{p-q}{p+q}$ is always negative and integer in this case, $z=-2,-3,-4,\ldots$, given that $p>q$ and $p+q =-2<0$.}
\label{fig:sphere}
\end{figure}

Let us now discuss the case of the sphere. We claim that the theory with $z=-2$ and $g=0$ is given by a single copy of $\doublewidetilde{D}_N$. It is straightforward to check that the central charges of the latter match \eqref{eq:apoly}, \eqref{eq:ccentral} with $z=-2$ and $g=0$. For $z<-2$ we have instead a linear quiver of $-z-3$ copies of $D_N$ with a copy of $\widetilde{D}_N$ at both ends. All the building blocks are coupled through $\mathcal{N}=2$ vector multiplets. Figure \ref{fig:sphere} shows the relevant examples. Evaluating the trial central charge is now straightforward: For $z=-3$ one can show that the trial $a$ and $c$ central charges are the same as those of $\doublewidetilde{D}_N$ theory plus \eqref{eq:newc}. Once again, this reproduces the anomaly polynomial computation. For $z<-3$ it is obvious that we just need to add the contribution \eqref{eq:newc} $-z-3$ times and, as we have already noticed, this modification just amounts to a shift in $z$. In particular, in the $g=0$ case adding \eqref{eq:newc} once decreases $z$ by one unit, which is precisely what we wanted.


\section{Heavy operators} 
\label{sec:heavy}

As is well known, the $T_N$ theory has a chiral operator $Q^{ijk}$ in the trifundamental of $\SU(N)_A\times \SU(N)_B \times \SU(N)_C$ with charge $(N-1)/2$ under $I_3$ and zero under $R_{\mathcal{N}=2}$.\footnote{\label{foot:Qtilde} $T_N$ actually contains a whole host of chiral operators $Q_{(k)}$, $k=1,\ldots,N-1$, in the $(\wedge^k,\wedge^k,\wedge^k)$ representation of $\SU(N)_A \times \SU(N)_B \times \SU(N)_C$, $\wedge^k$ being the $k$-index antisymmetric representation of $\SU(N)$ (we use the notation of \cite{maruyoshi-tachikawa-yan-yonekura}). $Q_{(k)}$ has R-charge $\frac{k}{2}(N-k)$ under $I_3$, and zero under $R_{\mathcal{N}=2}$. In particular $Q_{(1)}$ is the trifundamental operator of components $Q^{ijk}$ used throughout this paper, where the first index labels the fundamental of $\SU(N)_A$, the second that of $\SU(N)_B$, and the third that of $\SU(N)_C$.} After the nilpotent Higgsing which leads to $D_N$, this operator decomposes into $N$ operators in the bifundamental of the remaining $\SU(N) \times \SU(N)$ global symmetry. Their charges under the $\rho$ generator defined in \eqref{eq:gen} are:
\begin{equation}\label{eq:RQnew}
\rho(Q^{ijk})=\frac{N+1-2k}{2}\ ,\quad k=1,\dots,N\ .
\end{equation}
Their charges under $R_{\text{new}}$ are then:\footnote{As will be explained in section \ref{sub:Dn}, if we decide to close the $X=C$ puncture of $T_N$ to obtain $D_N$, each of the bifundamentals of the latter will be given by the components $Q_{T_N}^{ijk}$ for \emph{fixed} $k$. (This integer is the one labeling the fundamental of $\SU(N)_C$ of $T_N$.) This is the origin of the $k$ dependence in formulae \eqref{eq:RQnew} and \eqref{eq:RQDn}.}
\begin{equation}\label{eq:RQDn}
R_{\text{new}}(Q^{ijk})=\frac{1-\epsilon}{2}(N-1)-(1+\epsilon)\frac{N+1-2k}{2}\ .
\end{equation}

In a generalized quiver built out of $T_N$ theories, we can form a gauge invariant operator by multiplying all their trifundamentals 
together. This leads to a chiral operator which in the holographic setup is described by an M2-brane wrapped on the Riemann surface \cite{bbbw}. It is then natural to ask what is the field theory description of such ``heavy operator'' for inaccessible theories. Since to construct the latter we have to include $D_N$ building blocks in the generalized quiver, the most obvious guess would just be to consider the gauge invariant operator obtained by multiplying together the trifundamentals of $T_N$ blocks and the bifundamentals of the $D_N$ ones. However, as we have just seen, each $D_N$ block is equipped with $N$ bifundamentals of different R-charge and this leads to an ambiguity in the identification of \emph{the} heavy operator. In fact a generalized quiver with $n$ $D_N$ blocks has $N^n$ candidate heavy operators, and we should understand which one really corresponds to the M2-brane operator. 

As we have seen in section \ref{subsub:highgenus}, a theory associated with a geometry defined by line bundles of degree $p=2g-2+n$ and $q=-n$ can be constructed using $n$ $D_N$ blocks and $2g-2$ $T_N$ theories. The resulting heavy operators have R-charge
\begin{equation}
R_{\text{new}}(O_\text{heavy})=(1-\epsilon)(g-1)(N-1)-(1+\epsilon N)n+(1+\epsilon)\sum_{i=1}^{n}k_i\ .
\end{equation}
The parameters $k_i$ are integers and can take any value between 1 and $N$. 

The case of the sphere is even more complicated since we need to close multiple punctures in the same $T_N$ block. Every time we close a puncture we introduce a corresponding generator $\rho$ and indeed we should redefine the R-charge accordingly as explained in section \ref{sec:charges}. In the case of the $\widetilde{D}_N$ theory the trifundamental of $T_N$ decomposes into $N^2$ fundamentals of the residual $\SU(N)$ global symmetry. Their R-charges are:
\begin{equation}\label{eq:RQDtilden}
R_{\text{new}}(Q^{ijk})=\frac{1-\epsilon}{2}(N-1)-(1+\epsilon)(N+1-k_1-k_2)\ .
\end{equation} 
Again, $k_1$ and $k_2$ can take any value between 1 and $N$. 

Finally, the $\doublewidetilde{D}_N$ theory has no residual global symmetry, and the trifundamental of $T_N$ decomposes into $N^3$ chiral operators of charge:
\begin{equation}\label{eq:RQDtildetilden}
R_{\text{new}}(Q^{ijk})=\frac{1-\epsilon}{2}(N-1)-(1+\epsilon)\frac{3N+3-2k_1-2k_2-2k_3}{2}\ ,
\end{equation}
with $k_1,k_2,k_3=1,\ldots,N$.

The above analysis clearly shows that for inaccessible theories we always have multiple candidate heavy operators; however only one of them corresponds to the M2-brane operator we are looking for. As we will see in section \ref{sec:chiral}, these operators are not all independent due to nontrivial chiral ring relations. Moreover, we will give evidence that the M2-brane wrapping the Riemann surface corresponds to the operator obtained setting all $k_i$ parameters to one. Applying the recipe just advocated for we obtain the following R-charge for the heavy operator in all BBBW models:
\begin{equation}\label{eq:rheavy}
R(O_\text{heavy}) = \begin{cases}
\frac{(q-p)}{2}(N-1)\epsilon & \text{for}\ g=1 \\
(g-1)(N-1)(1-z\epsilon) & \text{for}\ g \neq 1
\end{cases}\ .
\end{equation}

Notice that this proposal (at leading order in $1/N$) reproduces precisely the holographic computation of \cite[Eq. 4.10]{bbbw}. We would like to stress once again that having at hand a precise field theoretic realization of these theories is what allowed us to determine the R-charge 
of the heavy operator exactly.


\section{Chiral ring relations} 
\label{sec:chiral}

We shall now see how the chiral ring relations for $T_N$ get modified by the addition of one, two, and three adjoint chiral fields $M_X$ ($X=A,B,C$), coupled to the former via the superpotential term $\Tr M_X \mu_X$. As is customary, we will refer to this process as ``flipping'' (or rotating) the puncture $X$ of $T_N$. Upon giving $M_X$ a nilpotent vev of the form \eqref{eq:vev} (i.e. completely ``closing'' the puncture), we obtain $D_N$, $\widetilde{D}_N$, and $\doublewidetilde{D}_N$, and the relations get further modified. 

\subsection{Summary of results}
\label{sub:chiral-results}

Let us summarize here the results of section \ref{sec:chiral}. These results represent the core of the paper. The reader not interested in their derivation can skip sections \ref{sub:Tn} through \ref{sub:Dntildetilde} and jump directly to section \ref{sec:bounds}.

\begin{itemize}
\item In $T_N$ we only consider the following relations among Higgs branch operators:
\begin{equation}
\Tr \mu_A^k = \Tr \mu_B^k = \Tr \mu_C^k\ , \quad k=2,\ldots,N\ ; \quad \mu_A Q = \mu_B Q = \mu_C Q\ \ .
\end{equation}
$Q$ is the trifundamental operator of $T_N$, while $\mu_X$ is the moment map associated with the $\SU(N)_X$ factor of the global symmetry of $T_N$.
\item In $T_N$ coupled to one, two, or three extra chirals $M_X$ we have (independently of their number):
\begin{equation}
\Tr \mu_A^k = \Tr \mu_B^k = \Tr \mu_C^k = 0\ , \quad k=2,\ldots,N\ ; \quad \mu_A Q = \mu_B Q = \mu_C Q = 0\ \ .
\end{equation}
Moreover, the moment map $\mu_X$ associated with the flavor symmetry $\SU(N)_X$ that is Higgsed by the nilpotent vev of $M_X$ is set to zero in the chiral ring:
\begin{equation}
\mu_X = 0\ .
\end{equation}
Finally we have:
\begin{equation}
M_A Q = M_B Q = M_C Q\ .
\end{equation}
When $M_C$ is not present we simply drop the last identity; when both $M_B$ and $M_C$ are not present, the above relation becomes $M_A Q = 0$.
\item In $D_N$ (obtained by closing the $X=A$ puncture of $T_N$) we have a single independent bifundamental operator $Q_{D_N}$, whose components are given by $Q_{T_N}^{1jk}$. $j$ and $k$ label the fundamental representations of the global symmetry groups which are not Higgsed, i.e. $\SU(N)_{B}$ and $\SU(N)_C$ respectively. We also have the relation:
\begin{equation}
\mu_B Q_{D_N} = \mu_C Q_{D_N}\ .
\end{equation}
Finally, there is a relation involving the bifundamental $Q_{D_N}$ and a particular combination of the nonzero components $a_l$ of the field $M_A$.
\item In $\widetilde{D}_N$ (obtained by closing the $X=A,B$ punctures) we have $N$ independent fundamental operators $Q^l_{\widetilde{D}_N}$, $l=1,\ldots,N$, whose components are given by $Q_{T_N}^{1lk}$ (or equivalently by $Q_{T_N}^{l1k}$). $k$ labels the fundamental representation of $\SU(N)_C$, which is not Higgsed. The nonzero components of $M_A$ and $M_B$ satisfy the simple relation $a_l = b_l$ for every $l=1,\ldots,N-1$. When $\widetilde{D}_N$ is coupled to an extra chiral $M_C$ (not given a nilpotent vev though), there remains a single independent (in the chiral ring) fundamental operator, given by $l=1$.
\item In $\doublewidetilde{D}_N$ (obtained by closing all three punctures of $T_N$) we have $N$ scalar operators $Q_{\doublewidetilde{D}_N} $ given by $Q^{11k}_{T_N}$, $k=1,\ldots,N$ (or equivalently by $Q_{T_N}^{1k1}$ or $Q_{T_N}^{k11}$). They are singlets under the full global symmetry of $T_N$ (i.e. $\SU(N)_A \times \SU(N)_B \times \SU(N)_C$ -- which is completely Higgsed). The nonzero components of $M_A$, $M_B$ and $M_C$ satisfy the simple relation $a_l = b_l = c_l$ for every $l=1,\ldots,N-1$. The scalar operator with the correct R-charge to match the energy of an M2-brane wrapping the Riemann surface is $Q^{111}_{T_N}$.
\end{itemize}

\subsection{\texorpdfstring{$T_N$}{TN}}
\label{sub:Tn}

The chiral ring relations for the $T_N$ theory\footnote{For a nice review on $T_N$ and its properties see \cite{tachikawa-TN}.} have been obtained in \cite{gaiotto-neitzke-tachikawa,benini-tachikawa-wecht,maruyoshi-tachikawa-yan-yonekura}. The ones we will be most interested in read:\footnote{As in \cite{maruyoshi-tachikawa-yan-yonekura}, we only focus on chiral ring relations among Higgs branch operators. Moreover, in the following sections we will be completely neglecting chiral ring relations involving the anti-trifundamental $Q_{(N-1)} = Q_{ijk}$ of $T_N$ (see footnote \ref{foot:Qtilde} for the notation). The modified relations holding in the new $D_N$, $\widetilde{D}_N$, $\doublewidetilde{D}_N$ blocks could be easily derived from those for $Q_{(1)}=Q^{ijk}$ by conveniently adapting our formulae.}
\begin{equation}\label{eq:relTn}
\Tr \mu_A^k = \Tr \mu_B^k = \Tr \mu_C^k\ , \ k=2,\ldots,N\ ; \qquad \mu_A Q = \mu_B Q = \mu_C Q\ \ .
\end{equation}
$Q=Q^{ijk}$ is the trifundamental of $T_N$, while $\mu_X$ ($X=A,B,C$) is the moment map associated with the $\SU(N)_X$ factor of the global symmetry of $T_N$.

\begin{figure}[ht]
\centering
\includegraphics{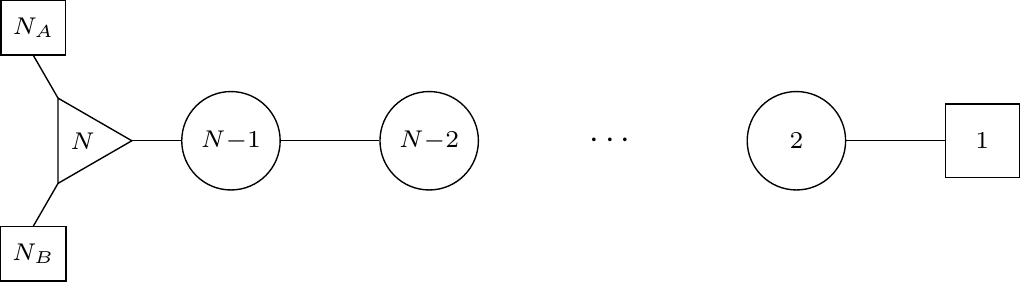}
\caption{Coupling a $T_N$ block to a linear quiver with $N-2$ gauge groups of decreasing rank. A subgroup $\SU(N-1)$ of $\SU(N)_C$ has been gauged by the first $\SU(N-1)$ gauge group on the right.}
\label{fig:TNcoupled}
\end{figure}

Let us reproduce here the original derivation presented in \cite{maruyoshi-tachikawa-yan-yonekura}, as very similar arguments will be used in later sections. Consider coupling $T_N$ to an $\SU$ linear quiver with decreasing ranks, starting off with $\SU(N-1)$. The latter gauges an $\SU(N-1)$ subgroup of, say, $\SU(N)_C$ of $T_N$. There are bifundamental hypermultiplets connecting the gauge groups (and an extra flavor of the last $\SU(2)$ factor); see figure \ref{fig:TNcoupled}.

\begin{figure}[ht]
\centering
\includegraphics{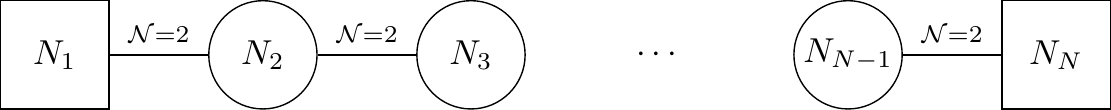}
\caption{A dual description of the linear quiver in figure \protect\ref{fig:TNcoupled}. $\SU(N)_1$ is identified with $\SU(N)_A$, whereas the special identification $\SU(N)_N \ni g \mapsto (g^{-1})^{t} \in \SU(N)_B$ exchanges fundamental and antifundamental representations of $\SU(N)_B$ and $\SU(N)_N$ \protect\cite{maruyoshi-tachikawa-yan-yonekura}.}
\label{fig:TNdual}
\end{figure}

\begin{figure}[h!t]
\centering

\begin{subfigure}[b]{0.9\textwidth}
\centering   
\includegraphics{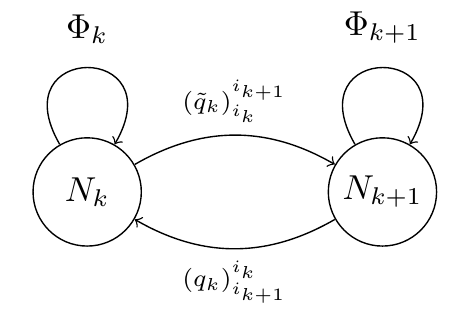}
\caption{The bifundamentals $q$ and $\tilde{q}$ between the $k$-th and $k+1$-th $\SU(N)$ gauge groups, $k=2,\ldots,N-1$. The upper (lower) index labels the (anti-)fundamental representation of the group with respect to which the arrow is incoming (outgoing). $\Phi_k$ is the $\mathcal{N}=1$ chiral multiplet (contained in the $\mathcal{N}=2$ vector multiplet) in the adjoint of the $\SU(N)_k$ gauge group.}
\label{fig:mu1N-a}
\end{subfigure}

\begin{subfigure}[b]{0.9\textwidth}
\centering
\includegraphics{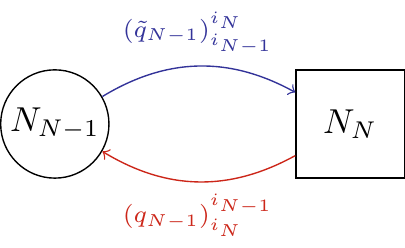}
\caption{The composition of arrows (from right to left) giving $\mu_N = ({\color{pigment3} \tilde{q}_{N-1}} {\color{link} q_{N-1}})_N = {\color{pigment3} \tilde{q}_{N-1}} {\color{link} q_{N-1}} - \frac{1}{N} \Tr {\color{pigment3} \tilde{q}_{N-1}} {\color{link} q_{N-1}}$, in the adjoint of the $N$-th group, i.e. the flavor $\SU(N)_N$.}
\label{fig:mu1N-b}
\end{subfigure}

\begin{subfigure}[b]{0.9\textwidth}
\centering
\includegraphics{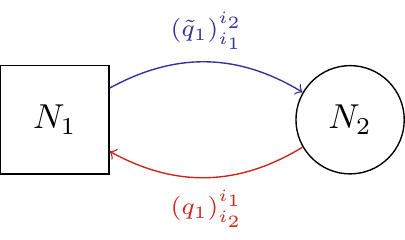}
\caption{The composition of arrows (from right to left) giving $\mu_1 = ({\color{link} q_{1}} {\color{pigment3} \tilde{q}_{1}} )_1 = {\color{link} q_{1}} {\color{pigment3} \tilde{q}_{1}}- \frac{1}{N} \Tr {\color{link} q_{1}} {\color{pigment3} \tilde{q}_{1}}$, in the adjoint of the first group, i.e. the flavor $\SU(N)_1$.}
\label{fig:mu1N-c}
\end{subfigure}
\caption{The bifundamentals appearing in the quiver of figure \protect\ref{fig:TNdual} and in equations \protect\eqref{eq:mus}.}
\label{fig:mu1N}
\end{figure}

This is dual \cite{gaiotto} to an $\SU(N)$ linear quiver with $N-2$ gauge nodes and two $\SU(N)$ flavor nodes at the tails, identified  with $\SU(N)_A$ and $\SU(N)_B$ of $T_N$; see figure \ref{fig:TNdual}. In this duality frame the adjoint chiral operators $\mu_{1,N}$ (i.e. the moment maps associated with $\SU(N)_{1,N}$) can be identified with the following ``mesons'':
\begin{equation}\label{eq:mus}
\mu_1 = (q_1 \tilde{q}_1)_1\ , \qquad \mu_N = (\tilde{q}_{N-1} q_{N-1})_N\ .
\end{equation}
Here a subscript $i$ on the bilinear means taking the adjoint with respect to the $i$-th group; see figures \ref{fig:mu1N-b} and \ref{fig:mu1N-c}.

\begin{figure}[ht]
\centering
\includegraphics{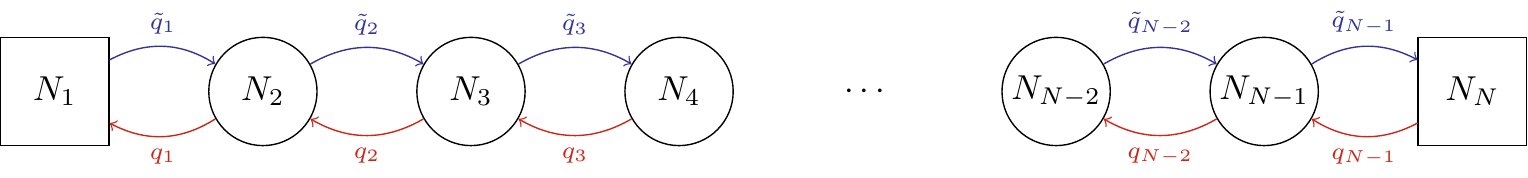}
\caption{The trifundamental $Q_{(1)}=Q^{i_{A}j_{B}k_{C}}$ of $T_N$ with $k_C=1$, i.e. ${\color{link} Q^{i_Aj_B1}}$, is given by the composition of arrows $({\color{link} q_1 q_2 \cdots q_{N-1}})^{\color{link} i_1}_{\color{link} j_N}$ (from right to left). The index $j_N$ labeling the antifundamental of $\SU(N)_N$ gets mapped to $j_{B}$, labeling the fundamental of $\SU(N)_B$; see figure \protect\ref{fig:TNdual}. (Equivalently, the anti-trifundamental is given by ${\color{pigment3} Q_{i_Aj_B1}} = ({\color{pigment3} \tilde{q}_{N-1} \tilde{q}_{N-2} \cdots \tilde{q}_{1}})^{\color{pigment3} j_N}_{\color{pigment3} i_1}$.)}
\label{fig:QTndual}
\end{figure}

Let us now focus on the  trifundamental operator $Q^{ijk}$ of $T_N$ ($i$ labels the fundamental of $\SU(N)_{A}$, $j$ that of $\SU(N)_{B}$, $k$ that of $\SU(N)_C$). To perform the gauging depicted at the left of figure \ref{fig:TNcoupled} we choose a subgroup $\SU(N-1) < \SU(N)_C$ spanning $k=2,\ldots,N$; $k=1$ is then a singlet ``direction''. In the dual quiver of figure \ref{fig:TNdual} when $k=1$ we can identify the operator $Q^{ij1}$ with the following ``baryon'' (i.e. chain of bifundamentals, see figure \ref{fig:QTndual}):
\begin{equation}\label{eq:Qs}
Q^{ij1} = (q_1 q_2 \cdots q_{N-1})^i_j\ . 
\end{equation}

We are now ready to prove \eqref{eq:relTn}. Consider the superpotential of the $\mathcal{N}=2$ linear quiver theory in figure \ref{fig:TNdual}:
\begin{equation}\label{eq:superpot}
\mathcal{W}_{\text{linear}} = \sum_{k=2}^{N-1} \Tr \left[ \Phi_k (-(\tilde{q}_{k-1}q_{k-1})_k+(q_k\tilde{q}_k)_k)\right]\ .
\end{equation}
The $\Phi_k$ F-terms (i.e. $\partial_{\Phi_k} \mathcal{W}_{\text{linear}}=0$) yield the relations
\begin{equation}\label{eq:eomq}
(\tilde{q}_{k-1}q_{k-1})_k = (q_k\tilde{q}_k)_k\ , \quad k=2,\ldots,N-2\ .
\end{equation}
\begin{figure}[ht]
\centering
\includegraphics{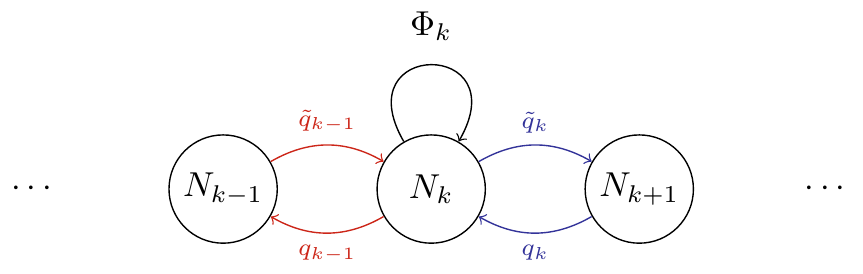}
\caption{Visualizing relation \protect\eqref{eq:eomq} for generic $k$ (i.e. the $\Phi_k$ F-term). The red ``petal'' ${\color{link} \tilde{q}_{k-1} q_{k-1}}$ equals the blue one ${\color{pigment3} q_{k} \tilde{q}_k}$. (The composition of arrows must be read from right to left.) Each petal gives a bilinear in the adjoint of the middle $\SU(N)_k$ gauge group.}
\label{fig:eomq}
\end{figure}
This immediately gives us
\begin{equation}\label{eq:eomtr}
\Tr \mu_1^k = \Tr \mu_N^k\ ,
\end{equation}
where we used cyclicity of the trace (which is not spoiled upon subtracting the traceless part). Clearly, the choice $X=C$ in figure \ref{fig:TNcoupled} is not special, and all of the above considerations can be repeated for any $\SU(N)_X$ factor of the flavor symmetry of $T_N$. We thus obtain the first in \eqref{eq:relTn}. The second is obtained as follows:
\begin{equation}\label{eq:rel2Tn}
(q_1 \tilde{q}_1)_1 q_1 \ldots q_{N-1} = q_1 (\tilde{q}_1 q_1)_2 q_2 \ldots q_ {N-1} = q_1 \ldots q_{N-1} (\tilde{q}_{N-1} q_{N-1})_N\ ,
\end{equation}
that is
\begin{equation}\label{eq:rel2Tnind}
(\mu_1)^i_l Q^{lj1} = (\mu_N)^j_l Q^{il1}\ .
\end{equation}
Given the full $\SU(N)_C$ symmetry of $T_N$, this equation must hold for any value of the third index $k$ of $Q^{ijk}$. Recalling that the choice $X=C$ in figure \ref{fig:TNcoupled} is arbitrary, we obtain the second relation in \eqref{eq:relTn}.

\subsection{\texorpdfstring{$T_N$ coupled to extra adjoint chirals}{TN coupled to extra adjoint chirals}}
\label{sub:TnM}

We will now couple $T_N$ to one, two, and three extra adjoint chirals $M_X$, $X=A,B,C$, and see how relations \eqref{eq:relTn} get modified. The coupling is specified by the superpotential term $\Tr M_X \mu_X$, where $\mu_X$ is the moment map associated with the $\SU(N)_X$ factor of the global symmetry of $T_N$.

\subsubsection{One chiral}
\label{subsub:1M}

\begin{figure}[ht]
\centering
\includegraphics{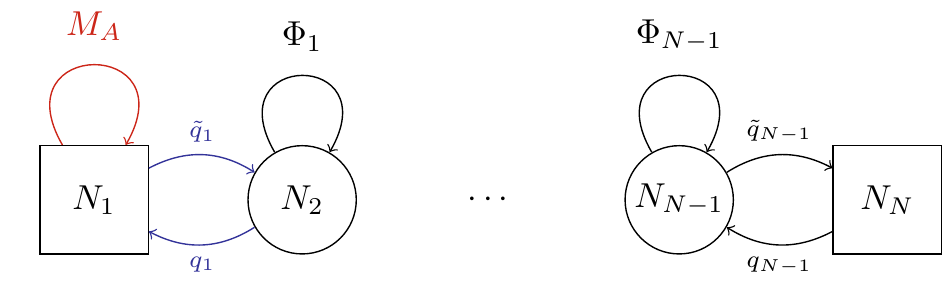}
\caption{Coupling the linear quiver of figure \protect\ref{fig:TNdual} to $M_A$ via $\Tr {\color{link} M_A} {\color{pigment3} \mu_1} = \Tr {\color{link} M_A} ({\color{pigment3} q_1\tilde{q}_1})_{\color{pigment3} 1}$.}
\label{fig:TN1M}
\end{figure}
\noindent We will exploit the dual linear quiver description of $T_N$ introduced in figure \protect\ref{fig:TNdual}. In this duality frame, let us assume that an extra adjoint chiral $M_A$ is coupled to $\mu_1$ (see figure \protect\ref{fig:TN1M}). The new superpotential is given by:
\begin{equation}\label{eq:superpotM}
\begin{split}
\mathcal{W} &= \mathcal{W}_\text{linear} + \mathcal{W}_\text{coupling}  \\ 
    &= \sum_{k=2}^{N-1} \Tr \left[ \Phi_k (-(\tilde{q}_{k-1}q_{k-1})_k+(q_k\tilde{q}_k)_k)\right] + \Tr M_A (q_1 \tilde{q}_1)_1 \ .
\end{split}
\end{equation}
Since in this duality frame $M_A$ is a fundamental field, we have new F-terms:
\begin{subequations}
\begin{align}
&\partial_{M_A} \mathcal{W} = \mu_1 = 0\ ; \label{eq:mu1Dn}\\
&\partial_{\tilde{q}_1} \mathcal{W} = M_A q_1 - q_1 \Phi_2 = 0\ ; \label{eq:q1}\\
&\partial_{\tilde{q}_{k}} \mathcal{W} = \Phi_k q_k - q_k \Phi_{k+1} = 0 \ , \quad k =2,\ldots,N-2\ ; \label{eq:qk} \\
&\partial_{\tilde{q}_{N-1}} \mathcal{W} = \Phi_{N-1} q_{N-1} = 0 \ . \label{eq:qN1}
\end{align}
\end{subequations}
First of all, \eqref{eq:mu1Dn} tells us that $0=\Tr \mu_1^k = \Tr \mu_N^k$ for any $k$, due to \eqref{eq:eomtr}. Recalling that the choice $X=C$ in figure \ref{fig:TNcoupled} is arbitrary, the first in \eqref{eq:relTn} then yields
\begin{equation}\
0 = \mu_A = \Tr \mu_A^k = \Tr \mu_B^k = \Tr \mu_C^k\ , \quad k=2,\ldots,N\ .
\end{equation}
The moment map associated with the puncture that has been flipped (due to the extra chiral $M_A$) is set to zero in the chiral ring. Furthermore, using the F-terms \eqref{eq:q1}, \eqref{eq:qk}, and \eqref{eq:qN1}, we obtain
\begin{equation}\label{eq:MQlong}
M_A q_1 q_2 \cdots q_{N-1} = q_1 \Phi_2 q_2 \cdots q_{N-1} = q_1 \cdots q_{N-2} \Phi_{N-1} q_{N-1} = 0\ ,
\end{equation}
which extends to 
\begin{equation}\label{eq:MQDn}
(M_A)^i_l \,Q^{ljk} = 0
\end{equation}
for any $k$ labeling the fundamental of $\SU(N)_C$ (a sum over $l$ is understood). 

We have seen the effect of $M_A$ on the first relation in \eqref{eq:relTn}; let us now see its effect on the second. Plugging \eqref{eq:mu1Dn} into the latter we immediately obtain 
\begin{equation}\label{eq:musTmM}
\mu_A Q = \mu_B Q = \mu_C Q=0\ .
\end{equation}
All in all, in $T_N$  coupled to $M_A$ we have the following relations in the chiral ring (all flavor and gauge index contractions are understood):
\begin{subequations}\label{eq:ringTnM}
\begin{align}
0 &= \mu_A \ ; \\
0 &= M_A Q \ ; \\
0 &= \Tr \mu_A^ k = \Tr \mu_B^k = \Tr \mu_C^k \ ,\quad k=2,\ldots,N\ ;\\
0 &= \mu_A Q = \mu_B Q = \mu_C Q \ . 
\end{align}
\end{subequations}

\subsubsection{Two chirals}
\label{subsub:2M}

Let us couple $T_N$ to two chiral fields $M_A$ and $M_B$. The superpotential of the linear quiver of figure \ref{fig:TNdual} coupled to the new fields becomes:
\begin{equation}\label{eq:WDntilde}
\mathcal{W}= \sum_{k=2}^{N-1} \Tr \left[ \Phi_k (-(\tilde{q}_{k-1}q_{k-1})_k+(q_k\tilde{q}_k)_k)\right] + \Tr M_A (q_1 \tilde{q}_1)_1 - \Tr M_B (\tilde{q}_{N-1} q_{N-1})_N\ .
\end{equation}
We have the following F-terms:
\begin{subequations}
\begin{align}
&\partial_{M_A} \mathcal{W} = \mu_1 = 0\ ; \label{eq:mu1Dntilde}\\
&\partial_{M_B} \mathcal{W} = \mu_N = 0\ ; \label{eq:muNDntilde}\\
&\partial_{\tilde{q}_1} \mathcal{W} = M_A q_1 - q_1 \Phi_2 = 0\ ; \label{eq:q1Dnt}\\
&\partial_{\tilde{q}_{k}} \mathcal{W} = \Phi_k q_k - q_k \Phi_{k+1} = 0 \ , \quad k =2,\ldots,N-2\ ; \label{eq:qkDnt} \\
&\partial_{\tilde{q}_{N-1}} \mathcal{W} = \Phi_{N-1} q_{N-1} -q_{N-1} M_B= 0 \ . \label{eq:qN1Dnt}
\end{align}
\end{subequations}
\noindent Using \eqref{eq:q1Dnt}, \eqref{eq:qkDnt}, and \eqref{eq:qN1Dnt} we obtain:
\begin{equation}\label{eq:MMQlong}
M_A q_1 q_2 \cdots q_{N-1} = q_1 \cdots q_{N-2} \Phi_{N-1} q_{N-1} =  q_1 \cdots q_{N-2} q_{N-1} M_B\ ,
\end{equation}
that is
\begin{equation}\label{eq:MQDnk1}
(M_A)^i_l \,Q^{lj1} = (M_B)^j_l \, Q^{il1}\ ,
\end{equation}
upon correctly identifying the indices of $\SU(N)_B$ with those of $\SU(N)_N$ as explained in the captions of figures \protect\ref{fig:TNdual} and \protect\ref{fig:QTndual}. The above equation must hold for any $k$, thus
\begin{equation}\label{eq:MQDntilde}
(M_A)^i_l \,Q^{ljk} =  (M_B)^j_l \, Q^{ilk}\ .
\end{equation}
All in all, in $T_N$ coupled to two chirals $M_A$ and $M_B$ we have the following relations in the chiral ring:
\begin{subequations}\label{eq:ringTn2M}
\begin{align}
&\mu_A = \mu_B = 0 \ ; \label{eq:mumuDntilde}\\
&M_A Q = M_B Q \ ; \\
&\Tr \mu_A^ k = \Tr \mu_B^k = \Tr \mu_C^k = 0\ ,\quad k=2,\ldots,N\ ;\\
&\mu_A Q = \mu_B Q = \mu_C Q = 0\ . \label{eq:mumuQDntilde}
\end{align}
\end{subequations}

\subsubsection{Three chirals}
\label{subsub:3M}

Let us finally flip all the punctures of $T_N$. To deal with this theory we cannot solely rely on the linear quiver description of figure \ref{fig:TNdual}, since in that picture only two out of three flavor symmetry factors are explicitly realized. We have to proceed in a smarter way, leveraging the availability of another duality frame describing the same theory. 
\begin{figure}[ht]
\centering
  \begin{minipage}{1\textwidth}
     \captionsetup{type=figure}
        \centering
        \subcaptionbox{$T_N$ with three flipped punctures. 
        \label{fig:Tn3M-a}}[.45\columnwidth]{
        \includegraphics{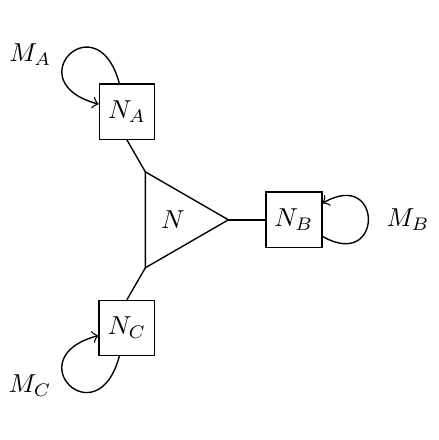}}\hspace{.5cm}
		\subcaptionbox{A different duality frame describing the same theory. 
        \label{fig:Tn3M-b}}[.45\columnwidth]{
        \vspace*{.8cm}
        \includegraphics[scale=.25]{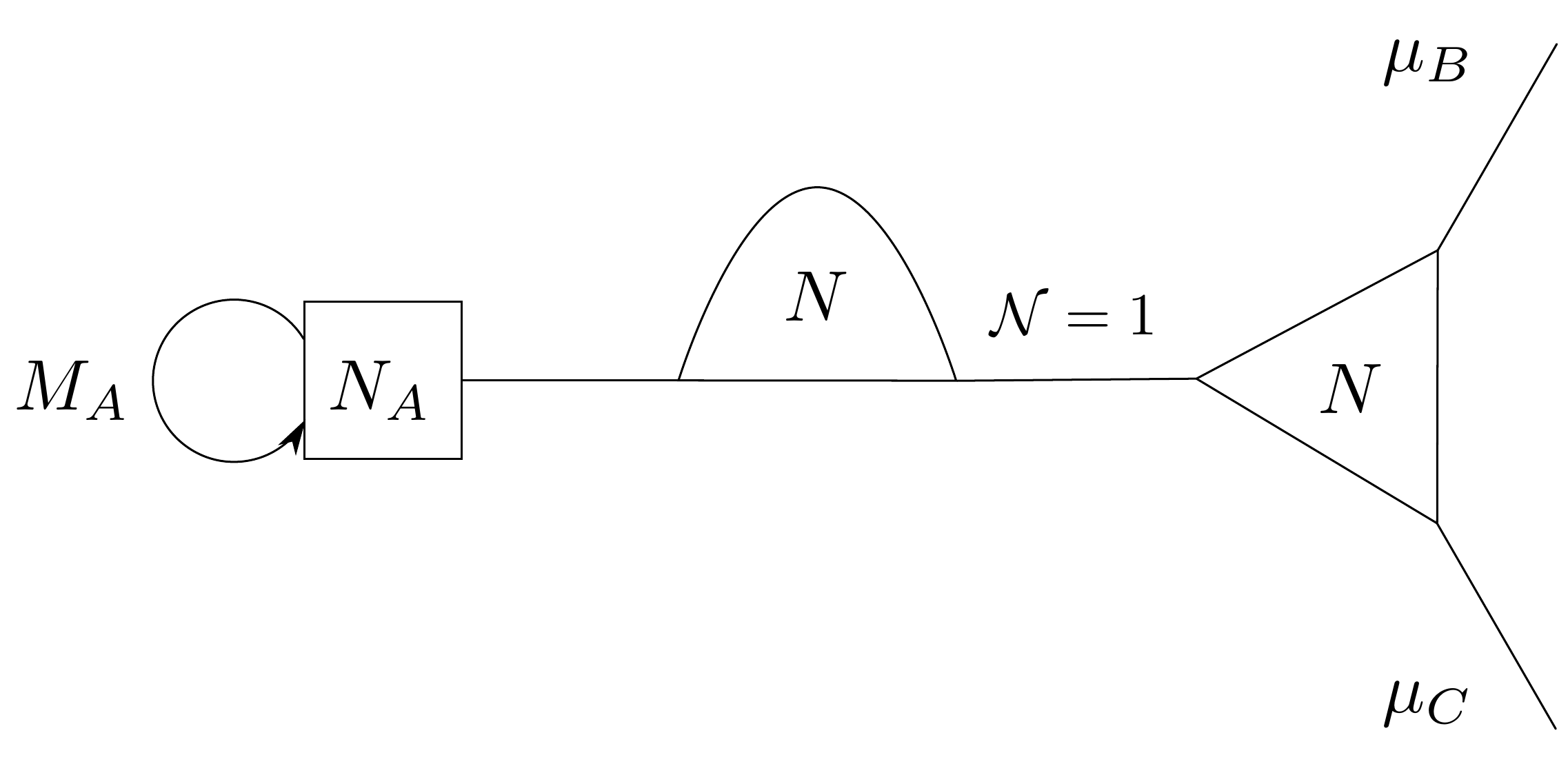}}
\end{minipage}
\caption{The equivalence of \protect\ref{fig:Tn3M-a} and \protect\ref{fig:Tn3M-b} can be derived starting from the duality called ``swap'' in \protect\cite{gadde-maruyoshi-tachikawa-yan} by closing a puncture in both duality frames. We will give a field theoretic proof of this statement in appendix \protect\ref{app:dualities}. In figure \protect\ref{fig:Tn3M-b}, the $\mathcal{N}=1$ gauging is described by the superpotential term $\Tr \mu_{D_N} \mu_X$, where $\mu_{D_N} $ is the moment map associated with the $\SU(N)$ global symmetry of the $D_N$ block connected to $T_N$ by a tube.}
\label{fig:Tn3Mdualframes}
\end{figure}
As explained in the introduction, this duality can be justified by matching the global symmetries available in the two frames (i.e. number of external legs), the structure of their superconformal indices, and their $a$ and $c$ central charges (which can be done explicitly by using the contribution of the modified tubes to $a$ given in section \ref{subsub:highgenus}). \newline

By inspecting figure \ref{fig:Tn3Mdualframes} we conclude that:\footnote{We will employ the notation $T_N^\mathrm{L,R}$ to mean the $T_N$ block in the left, respectively right, duality frame of figure \ref{fig:Tn3Mdualframes}.}
\begin{equation}
M_B Q_{T_N^\mathrm{L}} = Q_{D_N} Q_{T_N^\mathrm{R}} \mu_B\ , \qquad M_C Q_{T_N^\mathrm{L}} = Q_{D_N} Q_{T_N^\mathrm{R}} \mu_C\ .
\end{equation}
The contraction $Q_{D_N} Q_{T_N^\mathrm{R}}$ between the bifundamental of $D_N$ and the trifundamental of $T_N^\mathrm{R}$ (along the index $i$ labeling the fundamental of $\SU(N)_A$ that has been gauged) must produce the unique trifundamental of $T_N^\mathrm{L}$. Using the second in \eqref{eq:relTn}, valid in $T_N$ without extra chirals, we get:
\begin{equation}\label{eq:MAMBT}
M_B Q_{T_N^\mathrm{L}} = M_C Q_{T_N^\mathrm{L}}\ .
\end{equation}
Actually, the above argument can be repeated for any combination of two chirals $M_{X,Y}$ (due to the complete ``leg symmetry'' of figure \ref{fig:Tn3M-a}), leading to the nontrivial relation
\begin{equation}\label{eq:MQMQMQ}
M_A Q_{T_N} = M_B Q_{T_N} = M_C Q_{T_N}\ .
\end{equation}

One might wonder whether the identification $Q_{T_N^\mathrm{L}} = Q_{D_N} Q_{T_N^\mathrm{R}}$ is ``unique''. In fact, suppose we obtain $D_N$ by closing the $X=A$ puncture of $T_N$. In section \ref{sub:Dn} we will prove that $D_N$ contains in principle $N$ bifundamentals $Q_{D_N}^l$, $l=1,\ldots,N$; however, those with $l>1$ are not independent in the chiral ring, and are written in terms of $Q^1$ (which has R-charge given by \eqref{eq:RQDn} with $k=1$) and the quantum fluctuations of $M_A$ (which have R-charge given by \eqref{eq:RnewMcharge}). It is easy to see then that the only contraction matching the R-charge of the unique trifundamental of $T_N^\mathrm{L}$ is given by $Q^1_{D_N} Q_{T_N^\mathrm{R}}$.

Notice also that, since two chirals $M_{X,Y}$ in figure \ref{fig:Tn3M-a} always play the role of the $\mu_{X,Y}$ in figure \ref{fig:Tn3M-b} (for any combination of $X,Y=A,B,C$), the former will also satisfy the following relation (directly derived from the first in \eqref{eq:relTn}):
\begin{equation}\label{eq:trMABC}
\Tr M_A^k = \Tr M_B^k = \Tr M_C^k\ , \quad k=2,\ldots,N\ .
\end{equation}

Finally, despite the lack of a superpotential for $T_N$, the coupling of the latter to the extra chirals is still specified by superpotential terms of the form $\Tr M_X \mu_X$ by hypothesis. Therefore, the F-terms of 
\begin{equation}
\Tr M_A \mu_A + \Tr M_B \mu_B + \Tr M_C \mu_C
\end{equation}
with respect to the $M_X$'s (which have standard kinetic terms) yield the last nontrivial relation:
\begin{equation}\label{eq:Tn3Mallmu}
\mu_A = \mu_B = \mu_C = 0\ .
\end{equation}

All in all, in $T_N$ coupled to three chirals $M_X$, $X=A,B,C$, we have the following relations in the chiral ring:
\begin{subequations} \label{eq:ringTn3M}
\begin{align}
&\mu_A = \mu_B = \mu_C = 0 \ ; \\
&M_A Q = M_B Q = M_C Q\ .
\end{align}
\end{subequations}
The first also trivially implies $\Tr \mu_X^k = 0$ ($k=2,\ldots,N$) and $\mu_X Q = 0$ for every $X$.

\subsection{\texorpdfstring{$D_N$}{DN}}
\label{sub:Dn}

As explained in section \ref{sec:charges}, to obtain the $D_N$ theory we first have to couple $T_N$ to an extra adjoint chiral $M_A$, and then give the latter a nilpotent vev. Concretely, consider a vev of the form \eqref{eq:vev}. The multiplet (i.e. vev plus quantum fluctuations around it) takes the form \eqref{eq:mvev}. Let us slightly modify the notation (this will prove useful later on), and rename $a_{i-1}$ the components $M_{N,i}$ ($i=2,\ldots,N$) of $M_A$ appearing in that equation:
\begin{equation}\label{eq:Mvev}
M_A = \begin{bmatrix} 0 & 1 & 0 & 0 & \cdots \\ a_1 & 0 & 1 & 0 & \cdots \\ a_2 & a_1 & 0 & 1 & \cdots \\ a_3 & a_2 & a_1 & 0 & \cdots \\ \vdots & \vdots & \vdots & & \ddots \end{bmatrix}\ .
\end{equation}

Let us see how relations \eqref{eq:ringTnM} get further modified. Upon inserting \eqref{eq:Mvev} into \eqref{eq:MQDn}, we obtain:
\begin{equation}\label{eq:QrelDn}
Q^{i+1,jk} + \sum_{l=1}^{i-1}a_l Q^{i-l,jk} = 0\quad \text{for}\quad i=2,\ldots,N-1\ .
\end{equation}
For $i=1,N$ the above equation has to be understood as follows:
\begin{equation}\label{eq:QrelDn1N}
i=1:\quad Q^{2jk} = 0\ ; \qquad i=N:\quad \sum_{l=1}^{N-1}a_l Q^{N-l,jk} = 0\ .
\end{equation}
We already notice that the bifundamental operator $Q^{1jk}$ is left unconstrained by the system (in fact it will turn out to be the only independent one). On the opposite end, due to the particular form \eqref{eq:Mvev} of $M_A$, $Q^{2jk}$ is set to zero in the chiral ring for any $N$. Consider e.g. the $N=5$ case; \eqref{eq:QrelDn} and \eqref{eq:QrelDn1N} give:
\begin{subequations}\label{eq:QrelD5}
\begin{align}
&Q^{2jk}=0\ , \\ &Q^{3jk}=-a_1 Q^{1jk}\ , \\ &Q^{4jk}=-a_2 Q^{1jk}\ , \\ &Q^{5jk}=(a_1^2-a_3) Q^{1jk}\ , \\ &Q^{1jk}(-2a_1 a_2 +a_4) = 0\ . \label{eq:QrelD5fifth}
\end{align}
\end{subequations}
These relations determine all the components $Q^{ijk}$ with $i \neq 1$ (of the trifundamental $Q_{T_N}$) in terms of $Q^{1jk}$, which is instead left unconstrained by the system. \eqref{eq:QrelD5fifth} has to be interpreted as a relation in the chiral ring of $D_N$ among the nonzero components of $M_A$ and the independent bifundamental $Q^{1jk}$; for bigger values of $N$ it will still involve the latter and a more general expression in the components of $M_A$.

One could also wonder whether higher powers of $M_A$ give rise to new relations among the components $Q^{ijk}$, since
\begin{equation}
(M_A^p)^i_l Q^{ljk} = (M_A^{p-1})^i_s (M_A)^s_l Q^{ljk} = 0\ .
\end{equation}
As it turns out, these equations are fully redundant upon using the solutions to \eqref{eq:QrelDn} and \eqref{eq:QrelDn1N}.

In conclusion, in $D_N$ there is only one independent bifundamental operator $Q_{D_N}$ whose components are given by $Q^{1jk}_{T_N}$. Clearly, had we chosen to close the $X=B$ ($C$) puncture instead of $A$ its components would have been given by $Q^{i1k}$ ($Q^{ij1}$).\newline

\begin{figure}[!ht]
\centering
  \begin{minipage}{1\textwidth}
     \captionsetup{type=figure}
        \centering
        \subcaptionbox{$D_N$.
        \label{fig:Dndual-a}}[.45\columnwidth]{\vspace*{1.2cm}\includegraphics[scale=.2]{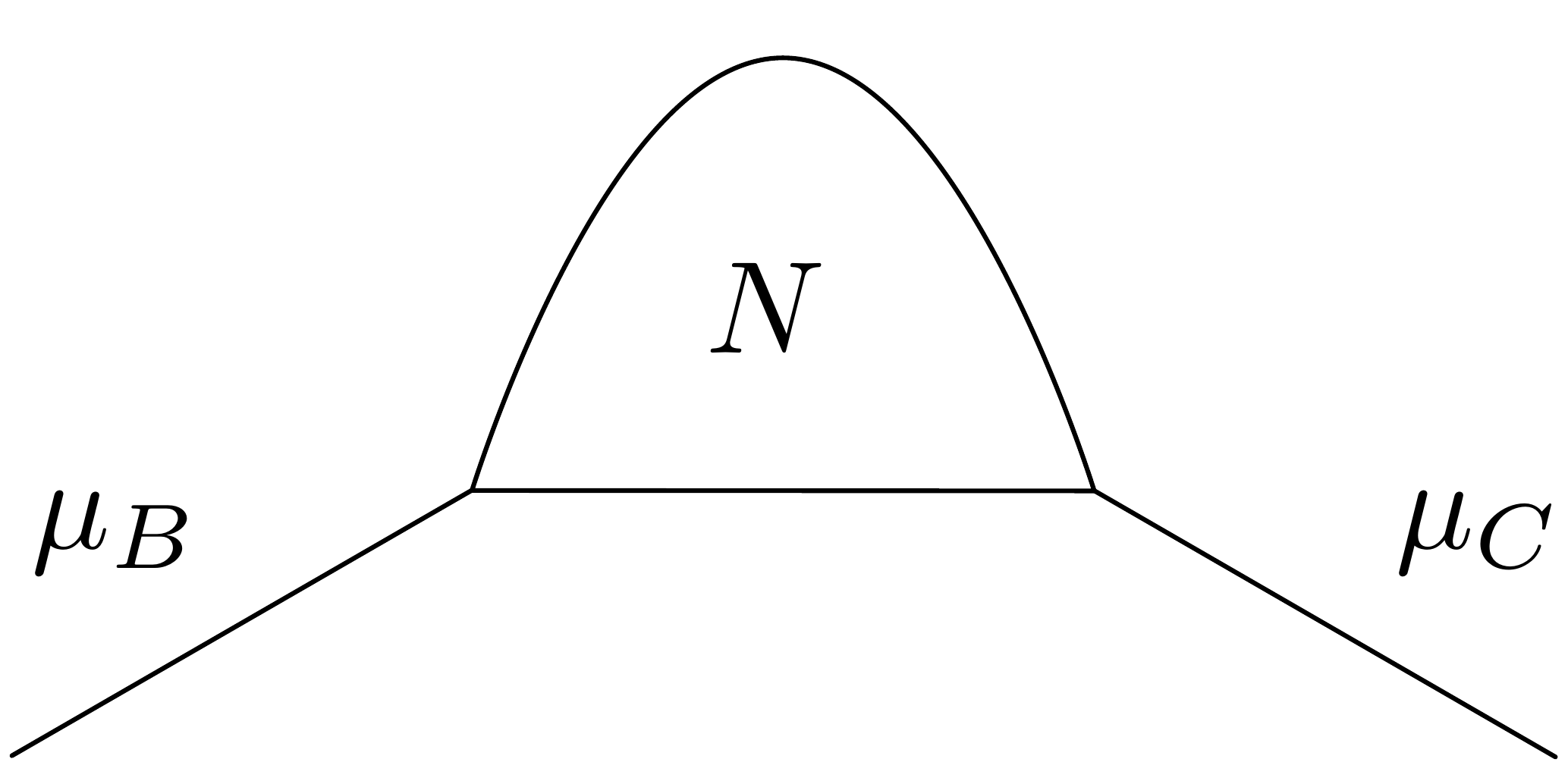}}\hspace{.5cm}
		\subcaptionbox{A different duality frame describing the same theory. 
        \label{fig:Dndual-b}}[.45\columnwidth]{\includegraphics[scale=.25]{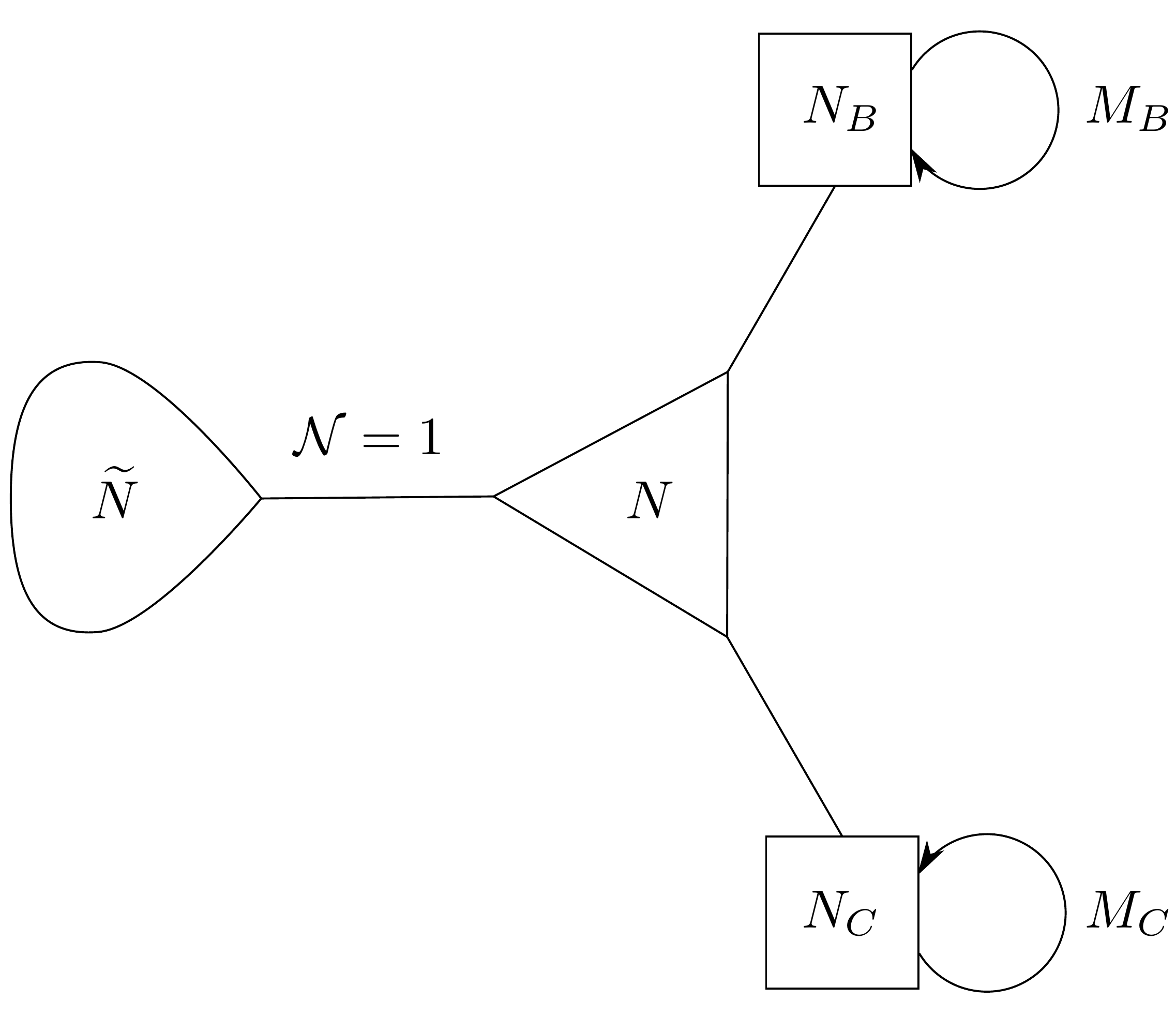}}
\end{minipage}
\caption{The equivalence of \protect\ref{fig:Dndual-a} and \protect\ref{fig:Dndual-b} is obtained from the duality depicted in figure \protect\ref{fig:Tn3Mdualframes} by closing a puncture in both frames (this will be proven in appendix \protect\ref{app:dualities}).}
\label{fig:Dndualframes}
\end{figure}

Let us now focus on relation \eqref{eq:musTmM}. That equation has been derived under the assumption that all the components of $M_A$ be generically nonzero. The nongeneric form \eqref{eq:Mvev} will now set some of the components of $\mu_A$ to zero, but will not impose any constraints on the others. We will employ a different duality frame to derive a nontrivial statement on the $\mu_X$'s of $D_N$. Consider then the two duality frames for $D_N$ depicted in figure \ref{fig:Dndualframes}.

Although we have not studied the properties of the $\widetilde{D}_N$ theory yet, at this level it is sufficient to know that it contains only one independent fundamental $Q_{\widetilde{D}}$, inherited from the trifundamental of $T_N$, much as $Q_{D_N}$ descends from $Q_{T_N}$. (We will prove this statement in section \ref{sub:Dntilde}.) Calling $\mu_{B,C}$ the moment maps associated with the two unflipped flavor symmetries of $D_N$, it suffices to inspect figure \ref{fig:Dndualframes} to conclude that
\begin{equation}\label{eq:DnmuAmuBno0}
\mu_B Q_{D_N} = Q_{\widetilde{D}_N} Q_{T_N} M_B \ , \qquad \mu_C Q_{D_N} = Q_{\widetilde{D}_N} Q_{T_N} M_C\ .
\end{equation}
The contraction $Q_{\widetilde{D}_N} Q_{T_N}$ between the fundamental of $\widetilde{D}_N$ and the trifundamental of $T_N$ (along the index $i$ labeling the fundamental of $\SU(N)_A$ that is being gauged) must produce the unique bifundamental of $D_N$ charged under $\SU(N)_B \times \SU(N)_C$. Using \eqref{eq:MQDntilde} in \eqref{eq:DnmuAmuBno0} gives:
\begin{equation}\label{eq:muAmuBDntilde}
\mu_B Q_{D_N} = \mu_C Q_{D_N}\ .
\end{equation}

\subsubsection{$D_N$ with extra chirals}
\label{subsub:Dnextra}

One could also couple $D_N$ to one or two extra chirals $M$, without giving the latter a nilpotent vev. This is the case when we use $D_N$ as building block in a generalized quiver, where the role of $M$ is played by the moment map of another theory (like $T_N$). In such a case, relations \eqref{eq:QrelDn} and \eqref{eq:QrelDn1N}, determining the ``higher'' bifundamentals $Q^l$, $l=2,\ldots,N$, of $D_N$ in terms of $Q^{1jk}_{T_N}$, become much more involved and contain the generic components of the extra chiral(s).

\subsection{\texorpdfstring{$\widetilde{D}_N$}{tildeDN}}
\label{sub:Dntilde}

To obtain the $\widetilde{D}_N$ theory we first have to couple $T_N$ to two extra adjoint chirals $M_{A,B}$, and then give the latter a nilpotent vev. According to our discussion at the beginning of section \ref{sub:Dn}, we will have:
\begin{equation}\label{eq:MABvevs}
M_A = \begin{bmatrix} 0 & 1 & 0   & \cdots \\ a_1 & 0 & 1  & \cdots \\ a_2 & a_1 & 0  & \cdots \\  \vdots & \vdots  & \vdots & \ddots \end{bmatrix}\ , \qquad M_B = \begin{bmatrix} 0 & 1 & 0   & \cdots \\ b_1 & 0 & 1  & \cdots \\ b_2 & b_1 & 0  & \cdots \\  \vdots & \vdots & \vdots & \ddots \end{bmatrix}\ .
\end{equation}
Inserting these expressions into \eqref{eq:MQDntilde} yields the set of equations:
\begin{equation}\label{eq:MQMQpost}
Q^{i+1,jk}-Q^{i,j+1,k}+\sum_{l=1}^{i-1}a_l Q^{i-l,jk} - \sum_{l=1}^{j-1}b_l Q^{i,j-l,k} = 0\ , \quad i,j=2,\ldots,N-1\ ;
\end{equation}
and for $i,j=1$ or $N$:
\begin{subequations}\label{eq:MQMQpost1N}
\begin{align}
&i=j=1 :  & &Q^{21k} - Q^{12k}=0 \ , \\
&i=1,\ j=N : & &Q^{2Nk}-\sum_{l=1}^{N-1}b_l Q^{1,N-l,k}= 0 \ , \\
&i=N,\ j=1 : & &\sum_{l=1}^{N-1}a_l Q^{N-l,1k} - Q^{N2k} = 0\ , \\
&i=j=N : & &\sum_{l=1}^{N-1}a_l Q^{N-l,Nk} - \sum_{l=1}^{N-1}b_l Q^{N,N-l,k}= 0 \ . 
\end{align}
\end{subequations}
The limiting $N=2$ case already displays the features of the generic one:
\begin{subequations}\label{eq:MQMQpost2}
\begin{align}
&Q^{21k} - Q^{12k} = 0\ , \\
&Q^{22k} - b_1 Q^{11k} = 0 \ , \\
&Q^{22k} - a_1 Q^{11k} = 0 \ , \\
&b_1 Q^{21k} - a_1 Q^{12k} = 0\ .
\end{align}
\end{subequations}
The above equations are solved by 
\begin{equation}
Q^{21k} = Q^{12k}\ , \quad a_1 = b_1\ , \quad Q^{22k} = a_1 Q^{11k}\ .
\end{equation}

Abstracting from the solution just given, we conclude that $\widetilde{D}_N$ contains $N$ independent fundamental operators $Q^j_{\widetilde{D}_N}$, $j=1,\ldots,N$ given by $Q^{1jk}_{T_N}$. Equivalently, they could be given by the components $Q^{i1k}_{T_N}$, since $Q^{(ij)k}$ is symmetric (as a tensor) in the two indices labeling the fundamental representations of the $\SU(N)_{A,B}$ flavor groups that have been Higgsed by the two chirals $M_{A,B}$. Also, the nonzero entries of $M_A$ and $M_B$ are forced to be equal: $a_l = b_l$ ($l=1,\ldots,N-1$).\newline

If we now flip the puncture of $\widetilde{D}_N$ (i.e. we couple it to $M_C$ without giving the latter a nilpotent vev to Higgs the $\SU(N)_C$ flavor group), we can further reduce the number of independent fundamental operators. We will use relation \eqref{eq:MQMQMQind} valid in $T_N$ coupled to three chirals $M_X$, $X=A,B,C$. The first identity in \eqref{eq:MQMQMQind} is nothing but \eqref{eq:MQDntilde}, which we know by now sets $a_l = b_l$ ($l=1,\ldots,N-1$) and implies that $Q^{(ij)k}$ be symmetric in the first two indices. We have to exploit the second identity in \eqref{eq:MQMQMQ}, that is $M_A Q = M_C Q$ or $M_B Q = M_C Q$: 
\begin{subequations}\label{eq:mixedDntilde}
\begin{align}
& Q^{i+1,jk} + \sum_{l=1}^{i-1}a_l Q^{i-l,jk} = (M_C)^k_m Q^{ijm}\ , & &i=2,\ldots,N-1\ , \\
& Q^{2jk} = (M_C)^k_m Q^{1jm}\ , & &i=1\ , \label{eq:mixed-1}\\
& \sum_{l=1}^{N-1}a_l Q^{N-l,jk} = (M_C)^k_m Q^{Njm}\ , & &i=N\ ; \\
& Q^{i,j+1,k} + \sum_{l=1}^{j-1}b_l Q^{i,j-l,k} = (M_C)^k_m Q^{ijm}\ , & &j=2,\ldots,N-1\ , \\
& Q^{i2k} = (M_C)^k_m Q^{i1m}\ , & &j=1\ , \label{eq:mixed-2}\\
& \sum_{l=1}^{N-1}b_l Q^{i,N-l,k} = (M_C)^k_m Q^{iNm}\ , & &j=N\ .
\end{align} 
\end{subequations}
Once again, suffice it to consider the $N=2$ case to see what kind of relations these equations are giving rise to. E.g. for $i=j=1$ we have:
\begin{equation}\label{eq:Dtilde1M}
Q^{21k} = (M_C)^k_m Q^{11m}\ , \quad Q^{12k} = (M_C)^k_m Q^{11m}\ .
\end{equation}
We see that $Q^{21k} = Q^{12k}$, as required by consistency with the relations for $\widetilde{D}_N$ (with unflipped puncture). Moreover, we conclude that for generic $N$ the ``higher'' fundamentals $Q^{1jk}$, $j>1$, can be written in terms of $Q^{11k}$. The relation will schematically read:
\begin{equation}\label{eq:Mcpoly}
Q^{1jk} = p_{j-1} \left( M_C \right) Q^{11k}\ , \quad j=2,\ldots,N\ ,
\end{equation}
where $p$ is a polynomial of degree $j-1$ in $M_C$, whose coefficients are proportional to the entries of $M_B$ (the $b_l$ in our notation).

Therefore in all instances in which $\widetilde{D}_N$ appears as a building block (possibly connected to other blocks like $T_N$ via a $\Tr M \mu$ interaction) the only independent fundamental operator $Q_{\widetilde{D}_N}$ is given by $Q^{11k}_{T_N}$.

\subsection{\texorpdfstring{$\doublewidetilde{D}_N$}{tildetildeDN}}
\label{sub:Dntildetilde}

We finally couple $T_N$ to three extra chirals $M_X$, $X=A,B,C$, and give all of them a nilpotent vev. We have:
\begin{equation}\label{eq:Mvevs}
 M_A = \begin{bmatrix} 0 & 1 & 0   & \cdots \\ a_1 & 0 & 1  & \cdots \\ a_2 & a_1 & 0  & \cdots \\  \vdots & \vdots  & \vdots & \ddots \end{bmatrix}\ , \quad M_B = \begin{bmatrix} 0 & 1 & 0   & \cdots \\ b_1 & 0 & 1  & \cdots \\ b_2 & b_1 & 0  & \cdots \\  \vdots & \vdots & \vdots & \ddots \end{bmatrix}\ , \quad M_C  = \begin{bmatrix} 0 & 1 & 0   & \cdots \\ c_1 & 0 & 1  & \cdots \\ c_2 & c_1 & 0  & \cdots \\  \vdots & \vdots & \vdots & \ddots \end{bmatrix}\ .
\end{equation}
We can make relation \eqref{eq:MQMQMQ} more explicit by writing:
\begin{equation}\label{eq:MQMQMQind}
(M_A)^i_l Q^{ljk} = (M_B)^j_l Q^{ilk} = (M_C)^k_l Q^{ijl}\ .
\end{equation}
In principle, we would now plug \eqref{eq:Mvevs} into \eqref{eq:MQMQMQind} and try to solve the equations in full generality. After the dust settles, we expect there to be only one independent scalar operator $Q_{\doublewidetilde{D}_N}$ (a singlet under the full global symmetry of $T_N$), given by a component of $Q_{T_N}$. As usual though, a general analysis would not be practical (nor enlightening). We tackle instead the simplest case, $N=2$, for which we just need to adapt \eqref{eq:QrelDn1N} to the present discussion. A quick inspection of the ensuing equations\footnote{We decided not to show them here as they are particularly cluttered.} reveals that $Q^{(ijk)}$ is totally symmetric (as a tensor) under the exchange $i \leftrightarrow j \leftrightarrow k$ (i.e. $Q^{211} = Q^{121} = Q^{112}$ and $Q^{122} = Q^{212} = Q^{221}$ for $N=2$), that the operator with two ``raised'' indices depends on the one with none (i.e. $Q^{122}$ depends on $Q^{111}$), and that the operator with all indices ``maximally raised'' depends on that with just one (i.e. $Q^{222}$ depends on $Q^{121}$). Finally, we also have $a_1 = b_1 = c_1$.\newline

All in all, for $N=2$ we get two independent scalar operators: $Q^{111}$ and $Q^{211}$. Similarly, for $N=3$ we get three, and so on, so that in the generic case we have $N$ independent scalar operators of the form $Q^{11k}$ ($k=1,\ldots,N$) as a result of the general analysis of \eqref{eq:MQMQMQind}. (The choice of the nontrivial index is unimportant in view of the complete symmetry of $Q^{(ijk)}$ as a tensor.) We also have $a_l = b_l = c_l$ for every $l=1,\ldots,N-1$.\newline

We could now ask whether there is some argument to select one among the $N$ scalar operators $Q^{11k}$ of $\doublewidetilde{D}_N$. A possible guiding principle would be to identify the one with the correct R-charge under $R_\text{new}$ in \eqref{eq:Repsnew}.  By ``correct'' we mean that, at least at leading order in $N$, it should match the energy (computed via holography) of an M2-brane wrapping the Riemann surface.\footnote{$\mathcal{C}_g$ is a sphere with three closed punctures in the case of $\doublewidetilde{D}_N$.} The R-charge of $Q^{11k}$ is given by formula \eqref{eq:RQDtildetilden} with $k_1=1$, $k_2=1$, and $k_3=k$. Our proposal (explained in section \ref{sec:heavy}) would be to select the operator with $k = k_3 = 1$. (We stress that those with ``raised'' $k$ index are also independent operators in the chiral ring.)

However, one should still worry about the $\mathcal{O}(1)$ differences between the R-charge of $Q^{111}$ and that of $Q^{11k}$ with $k \gtrapprox 1$ which are washed away in the large $N$ (i.e. holographic) limit.\footnote{On the contrary $Q^{11N}$ cannot be a good candidate, as its R-charge differs from that of $Q^{111}$ by an $\mathcal{O}(N)$ contribution, which we do not see from holography.} Thus, from this perspective the $k=k_3=1$ choice is not yet fully justified.

\begin{figure}[ht]
\centering
  \begin{minipage}{1\textwidth}
     \captionsetup{type=figure}
        \centering
        \subcaptionbox{$\doublewidetilde{D}_N$.
        \label{fig:Dnttdual-a}}[.2\columnwidth]{\includegraphics[scale=.085]{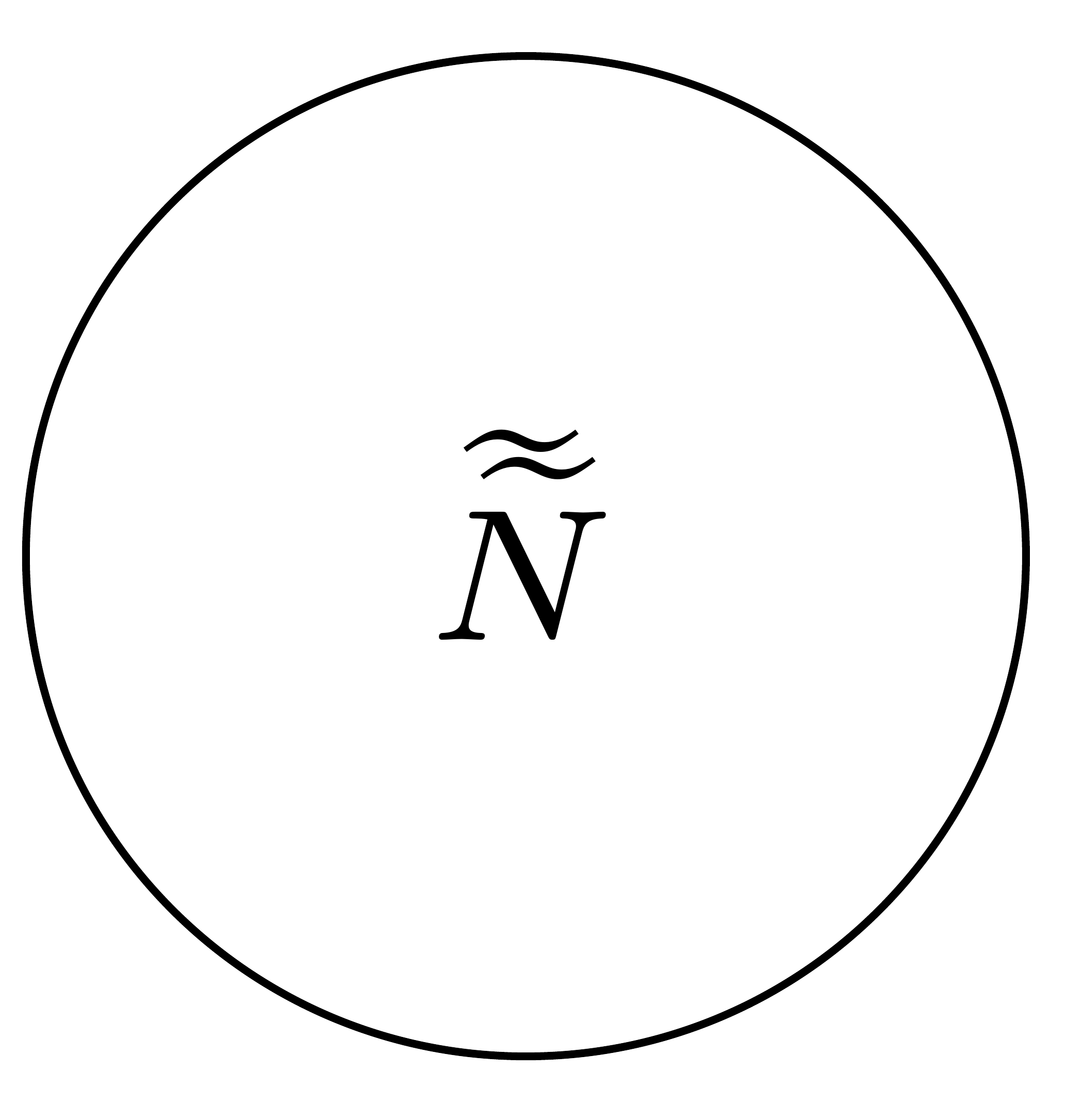}}\hspace{2cm}
		\subcaptionbox{A different duality frame describing the same theory. 
        \label{fig:Dnttdual-b}}[.45\columnwidth]{\includegraphics[scale=.25]{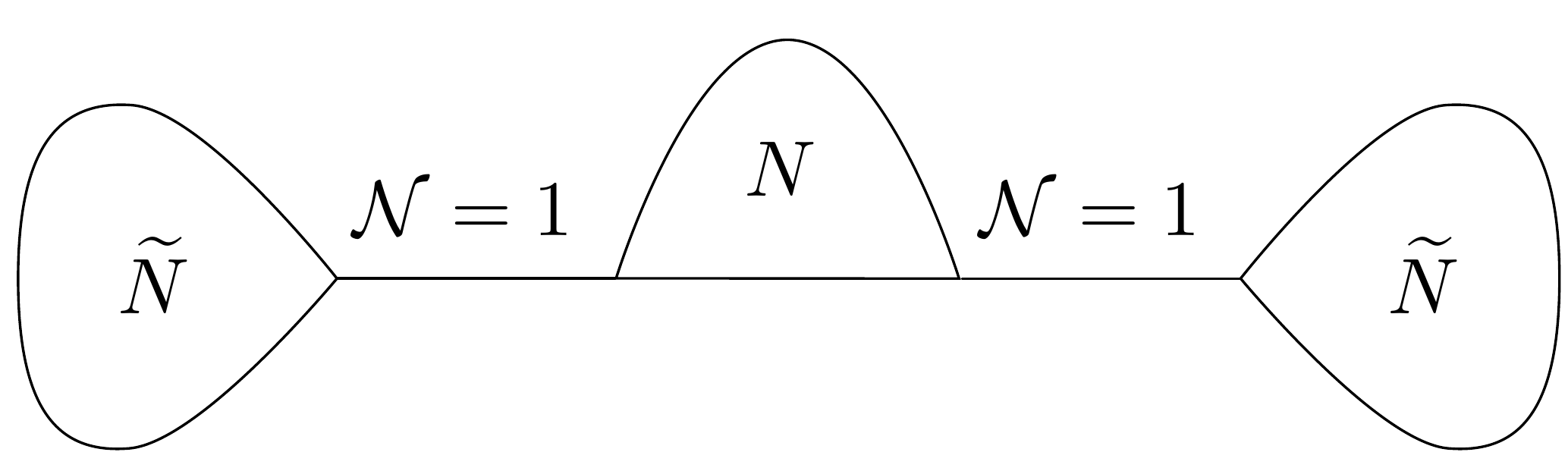}}
\end{minipage}
\caption{The equivalence of \protect\ref{fig:Dnttdual-a} and \protect\ref{fig:Dnttdual-b} descends from the duality depicted in figure \protect\ref{fig:Tn3Mdualframes} (this is proven in appendix \protect\ref{app:dualities}). In both duality frames there are no available flavor symmetries (i.e. no external legs). In figure \ref{fig:Dnttdual-b}, the $\mathcal{N}=1$ gaugings are specified by superpotential terms $\Tr \mu_{\widetilde{D}_N^\mathrm{L}} \mu_{D_N}$ and $\Tr \mu_{D_N} \mu_{\widetilde{D}_N^\mathrm{R}}$, where $\mu_{\widetilde{D}_N^\mathrm{L,R}}$ is the moment map associated with the unique $\SU(N)$ global symmetry of the left, right $\widetilde{D}_N$ block.}
\label{fig:Dnttdualframes}
\end{figure}

Luckily we can land further evidence, purely based on field theory, that this choice is the correct one. For this purpose, consider the two duality frames in figure \ref{fig:Dnttdualframes}, both describing $\doublewidetilde{D}_N$.

We want to match the correct scalar operator $Q_{\doublewidetilde{D}_N}$ with the singlet given by the contraction $Q_{\widetilde{D}_N^\mathrm{L}} Q_{D_N} Q_{\widetilde{D}_N^\mathrm{R}}$, of the form fundamental-bifundamental-fundamental. Since each of the blocks in figure \ref{fig:Dnttdual-b} contains a single (bi)fundamental operator, such a contraction must be unique, and has R-charge given by formula \eqref{eq:RQDtildetilden} with $k_1=k_2=k_3 =1$. This suggests that the right scalar operator in $\doublewidetilde{D}_N$ is indeed $Q^{111}_{T_N}$. Also, it is easy to see that those with $k=k_3 >1$ will correspond to a ``refinement'' of the contraction $Q_{\widetilde{D}_N^\mathrm{L}} Q_{D_N} Q_{\widetilde{D}_N^\mathrm{R}}$, i.e. to some longer string of operators of the form $Q_{\widetilde{D}_N^\mathrm{L}} \mu_{D_N}^k Q_{D_N} Q_{\widetilde{D}_N^\mathrm{R}}$. Let us see how. We can use the ``higher'' fundamentals $Q_{\widetilde{D}_N^\mathrm{L}}^l$ ($l>1$), which indeed have bigger R-charge. We know from the discussion below \eqref{eq:Mcpoly} that $Q_{\widetilde{D}_N^\mathrm{L}}^l = Q_{\widetilde{D}_N^\mathrm{L}}^1 p_{l-1}\left(M_{\widetilde{D}_N}\right)$, where the role of the extra chiral $M$ is here played by $\mu_{D_N}^\mathrm{L}$. For $l,m>1$, the most general contraction reads schematically:
\begin{equation}\label{eq:Dntt-string1}
Q_{\widetilde{D}_N^\mathrm{L}}^l Q_{D_N} Q_{\widetilde{D}_N^\mathrm{R}}^m \propto \sum_{p=1}^{l-1} \sum_{q=1}^{m-1} Q^1_{\widetilde{D}_N^\mathrm{L}} M_\mathrm{L}^p  \,Q_{D_N} M_\mathrm{R}^q \,Q^1_{\widetilde{D}_N^\mathrm{R}}\ ,
\end{equation}
with summands
\begin{equation}\label{eq:Dntt-string2}
Q^1_{\widetilde{D}_N^\mathrm{L}} M_\mathrm{L}^p\,  Q_{D_N} M_\mathrm{R}^q \,Q^1_{\widetilde{D}_N^\mathrm{R}} = Q_{\widetilde{D}_N^\mathrm{L}} \left(\mu^\mathrm{L}_{D_N}\right)^p Q_{D_N} \left(\mu^\mathrm{R}_{D_N}\right)^q Q_{\widetilde{D}_N^\mathrm{R}} = Q_{\widetilde{D}_N^\mathrm{L}} \left(\mu_{D_N}\right)^{p+q} \,Q_{D_N} Q_{\widetilde{D}_N^\mathrm{R}}\ .
\end{equation}
In the last identity we have used \eqref{eq:muAmuBDntilde} on $\mu^\text{L,R}_{D_N}$. As is obvious from this discussion, scalar operators in $\doublewidetilde{D}_N$ with $k=k_3 >1$ correspond to ``refined'' contractions in figure \ref{fig:Dnttdual-b}. These involve more operators, and have higher R-charge.


\section{Unitarity bound violations} 
\label{sec:bounds}

In \cite{bbbw} it was found that all models (i.e. both accessible and inaccessible theories) satisfy the Hofman--Maldacena bound \cite{hofman-maldacena,kulaxizi-parnachev} on $\frac{a}{c}$ with one exception: The theory on the sphere with $N=-z=2$, i.e. $\doublewidetilde{D}_2$ in our notation. From our perspective this model indeed has something special. It is the only lagrangian theory without any gauge interaction, as it is defined by a collection of chiral multiplets with quadratic and cubic superpotential terms. As was noticed in \cite{maruyoshi-song-def}, such a theory is free in four dimensions. The explanation of the apparent unitarity bound violation is that there are emergent $\U(1)$ symmetries which are not taken into account in the $a$-maximization procedure. Having given an explanation for such exception, we will discard this model in the rest of the discussion.

\subsection{\texorpdfstring{$M_{N,i}$ operators}{MNi operators}}
\label{sub:MNi}

As we have explained before, when we close a puncture $X$ the chiral ring of the resulting theory includes operators $M_{N,i}$ 
of R-charge $i(1+\epsilon)$ with $2\leq i\leq N$ coming from $M_X$ (called $a_{i-1}$, $b_{i-1}$ or $c_{i-1}$ in section \ref{sec:chiral}, for $X=A,B,C$ respectively). The value of $\epsilon$ depends on the theory at hand and is determined by $a$-maximization. In particular, the operators $M_{N,i}$ will violate the unitarity bound whenever $\epsilon<\frac{2-3i}{3i}$. 

It turns out that for all the theories associated with Riemann surfaces of genus greater than zero the value of $\epsilon$ which maximizes (\ref{eq:apoly}) is always greater than $-2/3$ and consequently all the operators $M_{N,i}$ are above the unitarity bound in these cases. The story is different for the sphere, and indeed we find an infinite family of models in which the $M_{N,2}$ operators do violate the unitarity bound: Such models are the theories with $z=-2$ and arbitrary $N$ (as has also been noticed in \cite{maruyoshi-song-def}), and those with $z=-3$ and $N$ greater than two. For $z<-3$ we always find $\epsilon>-2/3$, hence there are no unitarity bound violations. For the former cases we propose the standard interpretation whereby $M_{N,2}$ fields decouple and become free. Accordingly, we should subtract the contribution of the decoupled chiral multiplets to the trial $a$ central charge and repeat the $a$-maximization procedure. 

Following our dictionary, the theory on the sphere with $z=-2$ contains a single $\doublewidetilde{D}_N$ block whereas the theory with $z=-3$ is given by two copies of $\widetilde{D}_N$ coupled through an $\mathcal{N}=2$ $\SU(N)$ vector multiplet. Consequently, in the case $z=-2$ we have to subtract the contribution of the three chiral multiplets $M_{N,2}$, whereas in the case $z=-3$ there are five multiplets which decouple: The four $M_{N,2}$ and the gauge invariant combination $\Tr \Phi^2$ of the chiral multiplet contained in the $\mathcal{N}=2$ vector multiplet, which also has trial R-charge $2+2\epsilon$. We find that after the new maximization all the surviving operators satisfy the unitarity bound and the ratio $\frac{a}{c}$ satisfies the Hofman--Maldacena bound. 

\subsection{Heavy operators}
\label{sub:heavy}

We find that the heavy operators are always above the unitarity bound with one exception: The theory with $g=1$, $p=1$ and 
$N=2$. This theory is described by a single $D_2$ block with the diagonal combination of the two $\SU(2)$ global symmetry 
groups being gauged (via an $\mathcal{N}=2$ vector multiplet as usual); see figure \ref{fig:torus-1}. As in the previous section, the resolution is that the heavy operator decouples from the theory. Again, there is an emergent $\U(1)$ symmetry acting on this field which our procedure does 
not detect. 

Actually we can do more, and check field theoretically that this scenario is correct: This theory is just $\SU(2)$ $\mathcal{N}=4$ SYM plus a free hypermultiplet (the gauge invariant part of the trifundamental of $T_2$), with an extra chiral multiplet $M$ in the adjoint of the $\SU(2)$ symmetry group of the theory. The coupling between the $\SU(2)$ moment map and $M$ further breaks extended supersymmetry down to $\mathcal{N}=1$. When we give a nilpotent vev to $M$, one of the decoupled chiral fields becomes massive and we are left with a single decoupled chiral which is precisely our heavy operator. 

The rest of the theory is $\SU(2)$ SYM with three chiral multiplets in the adjoint and an extra singlet with superpotential given by
\begin{equation}
\mathcal{W}=\Tr(\Phi[X,Y])+Z\Tr Y^2-\Tr X^2\ .
\end{equation}
As is clear from the above formula, the multiplet $X$ becomes massive and when we integrate it out we are left with 
\begin{equation}
\mathcal{W}=\frac{1}{4}\Tr([\Phi,Y]^2)+Z\Tr Y^2\ .
\end{equation}
Assigning trial R-charge $1+\epsilon$ to $\Phi$, marginality of the superpotential assigns charge $-\epsilon$ to $Y$ 
(which is also imposed by R-symmetry anomaly cancellation) and $2+2\epsilon$ to $Z$. It is straightforward to check that the trial $a$ and $c$ central charges of this theory match those of the theory with $p=g=1$ and $N=2$, once we remove the contribution of the decoupled chiral of charge $-\epsilon$. 
The value of $\epsilon$ found by $a$-maximization is $\approx -0.54$, hence we do not find any further unitarity bound violation in this model. 


\section{Counting relevant operators} 
\label{sec:index}

By adapting the construction of \cite{gaiotto-rastelli-razamat}, the superconformal index of both accessible ($g>1$, $\vert z\vert\leq1$) 
and inaccessible theories has been proposed in \cite{beem-gadde}. By considering a special limit of the index, the authors of \cite{beem-gadde} have argued that the number of operators of charge $2+2\epsilon$ (in our convention) is always $g-1+\vert p-q\vert$. As has already been discussed in \cite{bbbw}, this class of operators is known to exhaust the list of relevant ones for accessible theories. Moreover this result leads to a precise prediction for the inaccessible theories as well.  
As we will now explain, our models pass this test. However, contrary to the case of accessible theories, for most of our models the class of operators proposed by \cite{beem-gadde} \emph{does not} exhaust the list of relevant operators.\newline
 
Motivated by the above considerations, in this section we will compute the number of relevant operators in our models. By \emph{relevant} we mean independent gauge invariant chiral multiplets whose R-charge is strictly smaller than two (hence we will be ignoring products of gauge invariant operators). For simplicity, let us start by neglecting the contribution from heavy operators, which indeed might be relevant for $N$ small enough.
For accessible theories it is known that the number of relevant operators is $g-1+\vert p-q\vert$. The only operators contributing to this number are the $\SU(N)$ moment maps of $T_N$ theories and the quadratic Casimirs built out of the $\mathcal{N}=1$ adjoint chirals inside the $\mathcal{N}=2$ vector multiplets. These operators all have R-charge $2+2\epsilon$ (for $z>0$). For $\vert z\vert\geq1$ there are also $D_{N}$ building blocks, hence the operators $M_{N,2}$ (which also have R-charge $2+2\epsilon$) contribute to the counting. Overall, the quadratic $\SU(N)$ Casimirs $\Tr\Phi^2$ plus the $M_{N,2}$ operators reproduce the above counting. However, this is not the end of the story. Since the maximum of the $a$ central charge is attained for $\epsilon<-1/3$, the operators with R-charge $3+3\epsilon$ (such as $M_{N,3}$ and $\Tr\Phi^3$) also contribute to the counting, leading to a total of $2g-2+2\vert p-q\vert$ operators. This is the result whenever $-1/3>\epsilon>-1/2$, whereas this counting has to be modified if $\epsilon<-1/2$, since $\Tr\Phi^4$ and $M_{N,4}$ also contribute in this range. In conclusion, we find the answer:
\begin{equation}\label{eq:gencount}
\#\ \text{relevant} = k(g-1+\vert p-q\vert)\quad\text{for}\quad -\frac{k-1}{k+1}>\epsilon>-\frac{k}{k+2}\quad\text{and}\quad N> k\ .
\end{equation}
If $N<k$ the result is instead $(N-1)(g-1+\vert p-q\vert)$, since there are neither $\Tr\Phi^k$ nor $M_{N,k}$ operators.

We will now discuss separately the cases $g>1$, $g=1$ and $g=0$ including heavy operators in the counting. For $g=0$ there is an additional subtlety we should take into account. So far we have neglected the components of the $\SU(N)$ 
moment map of $T_N$ coupled to the fields $M_{N,i}$, since F-terms set them to zero in the chiral ring. However, as we have 
already seen, for $z=-2$ and $z=-3$ the multiplets $M_{N,2}$ violate the unitarity bound and decouple. We then conclude that the 
corresponding components of the moment map, whose R-charge is $-2\epsilon$, are no longer zero in the chiral ring and should be 
included in the counting since they are gauge invariant. We have three such multiplets for $z=-2$, and four for $z=-3$.

\subsection{High genus}
\label{sub:g>1}

The value of $\epsilon$ which maximizes the $a$ central charge was computed in \cite[Eq. (2.17)]{bbbw}: 
\begin{equation}
\epsilon=\frac{N+N^2-\sqrt{z^2+(N+N^2)(N+N^2+z^2(4+3N+3N^2))}}{3z(1+N+N^2)}\ .
\end{equation}
At fixed $N$, $\epsilon$ is a monotonically decreasing function of $z$ and lies in the range $[-\frac{1}{2},-\frac{3}{5}]$ for $z\geq z_\mathrm{c}$, where 
\begin{equation}
z_\mathrm{c}(N) \equiv \frac{12(N+N^2)}{3N^2+3N-5}\ .
\end{equation}
From \eqref{eq:rheavy} and the above formula it is clear that the heavy operator is relevant only for $g=z=N=2$. Its R-charge is 
$1-2\epsilon$, which in this case is $\approx 1.85$. 
In conclusion, 
\begin{equation}\label{eq:rel-high-genus}
\# \ \text{relevant} = \begin{cases} 
6 & \text{for}\ N=2\ \text{and}\ z=g=2 \\
g-1+\vert p-q\vert & \text{for}\ N=2\ \text{and}\ z,g \neq 2 \\
2(g-1+\vert p-q\vert) & \text{for}\ N=3\ \text{and generic $z$, $g$} \\
2(g-1+\vert p-q\vert) & \text{for}\ N>3,\ z\leq z_\mathrm{c}\ \text{and generic $g$} \\
3(g-1+\vert p-q\vert) & \text{for}\ N>3,\ z > z_\mathrm{c}\ \text{and generic $g$}
\end{cases}\ .
\end{equation}

\subsection{Torus}
\label{sub:g=1}

In this case the value of $\epsilon$ which maximizes the $a$ central charge is independent of $p$: 
\begin{equation}
\epsilon=-\sqrt{\frac{3N^2+3N+1}{9N^2+9N+9}}\ .
\end{equation}
This formula is valid for every value of $N$ and $p$, except for $N=2$, $p=1$ due to the unitarity bound violation of the heavy 
operator. (Having already discussed this model in detail in section \ref{sub:heavy} we will not consider it here.) 
From the above formula we see that $\epsilon$ is always in the range $[-0.58,-0.54]$. This immediately implies that (neglecting 
heavy operators) the number of relevant operators is $\vert p-q\vert=2p$ for $N=2$,\footnote{In the case of the torus the twist parameter \eqref{eq:z} is not well-defined, and we revert to using $p$ to label the theories, since $2g-2=0=p+q \Leftrightarrow p=-q$ and $p>q$ by hypothesis.} $4p$ for $N=3$ and $6p$ for $N>3$. 

From \eqref{eq:rheavy} we see that the heavy operator is relevant for $N=2$ and $p=2,3$. In the case $p=2$ the heavy operator is 
just the product of the two $\SU(2)\times \SU(2)$ bifundamentals ($(q_1)_i^j$ and $(q_2)_i^j$) contained in $D_2$. Since 
$\epsilon \approx -0.55$, also $\Tr(q_1q_2\Phi_1)$ and $\Tr(q_2q_1\Phi_2)$ are relevant as their R-charge is $1-\epsilon$. (As usual the $\Phi_i$ denote the $\mathcal{N}=1$ chiral multiplets in the two $\mathcal{N}=2$ vector multiplets.) In a sense, we can ``dress'' the heavy operator 
by adding a $\Phi$ multiplet. Instead, we do not get any relevant operators other than the heavy operator in the case $p=3$. In conclusion we find seven relevant operators both for $p=2$ and $p=3$. Of course for larger values of $p$ the answer is $2p$.

In the case $N=3$ the heavy operator is relevant only for $p=1$ and its R-charge is $-2\epsilon$. Also in this case the heavy operator 
``dressed'' by the $\Phi$ multiplet is relevant and has charge $1-\epsilon$ . In conclusion we find six relevant operators in this 
case. Finally, for $N=4$ and $p=1$ only the heavy operator is relevant, leading to a total of seven. 

The results are neatly summarized in the following table.
\begin{table}[!htb]
\centering
{\renewcommand{\arraystretch}{1.2}%
\begin{tabular}{@{}l  c  c  c  c  c@{}}
& $N=2$ & $N=3$ & $N=4$ & $N>4$  \\
\toprule
$p = 1$  & \multicolumn{1}{!{\vrule width 1pt}c}{one $D_2$ block (section \ref{sub:heavy})} & 6 & 7 & 6 \\

$p= 2,3$ & \multicolumn{1}{!{\vrule width 1pt}c}{7} & \multicolumn{1}{c|}{$4p$} & \multicolumn{2}{c}{$6p$}\\

$p> 3$ & \multicolumn{1}{!{\vrule width 1pt}c}{$2p$} & \multicolumn{1}{c|}{$4p$} & \multicolumn{2}{c}{$6p$}\\
\bottomrule
\end{tabular}}
\caption{Total number of relevant operators (including the heavy ones) in the $g=1$ case.}
\label{tab:rel-torus}
\end{table}

\subsection{Sphere}

Finally, in the case of the sphere the value of $\epsilon$ which maximizes $a$ is:\footnote{Recall that $z<0$ for $g=0$, as explained in the caption to figure \ref{fig:sphere}.}
\begin{equation}\label{eq:parmax}
\epsilon=\frac{N+N^2+\sqrt{z^2+(N+N^2)(N+N^2+z^2(4+3N+3N^2))}}{3z(1+N+N^2)}\ .
\end{equation}
$\epsilon$ is a monotonically decreasing function of $z$ at fixed $N$. As we have noticed before, due to unitarity bound violations this formula is reliable only for $z<-3$. 

For $z=-2$ we instead find that
\begin{equation}
\epsilon_{z=-2}=\frac{36+N-N^3-\sqrt{100 + 176 N + N^2 - 136 N^3 - 10 N^4 + 13 N^6}}{6N^3-78}\ ,
\end{equation}
which lies in the interval $[-7/9,-3/4]$ for every $N\geq3$. From the previous discussion we thus conclude that (neglecting heavy 
operators) the number of relevant operators is $7(g-1+\vert p-q\vert)+3=24$ for $N\geq8$ and $3N$ for smaller 
values of $N$. This counting includes the three free chiral multiplets. 

For $z=-3$ we have:
\begin{equation}
\epsilon_{z=-3}=\frac{60+N-N^3-\sqrt{289 + 378 N + N^2 - 276 N^3 - 20 N^4 + 28 N^6}}{9N^3-129}\ ,
\end{equation}
which lies in the range $[-5/7,-2/3]$, leading to a total of $-5(1+2z)+4=29$ relevant operators for $N\geq6$ and $5N-1$ for 
$N=3,4,5$. Again this number includes the five free chiral multiplets. For $N=2$ there is no violation of the unitarity bound 
and \eqref{eq:parmax} holds. In this case we find five relevant operators and none of them is free.

From \eqref{eq:parmax} we find for $z<-3$ the inequalities $-2/3<\epsilon<-1/2$. Moreover, $\epsilon>-3/5$ for
\begin{equation}
\vert z\vert \geq z_\mathrm{c}(N) \equiv \frac{45 (N + N^2)}{28 + 3N + 3N^2}\ .
\end{equation}
As a result, the number of relevant operators for $N>4$ is $-4(1+2z)$ or $-3(1+2z)$ depending on whether $\vert z\vert$ is smaller 
or larger than $z_\mathrm{c}$. In the cases $N=2,3,4$ we find $-(N-1)(1+2z)$ for any value of $z$.

So far we have neglected possible heavy operators, to which we now turn our attention. 
For $N=2$ the heavy operator is relevant for $z\geq-4$. It has R-charge $\approx 0.96$ and $\approx 1.50$ for $z=-3$ and $z=-4$ respectively. 
For $z=-3$ we can obtain other relevant operators by dressing the heavy operator with $\Phi_{\SU(2)}$ or $\Phi_{\SU(2)}^2$. In the 
$z=-4$ case we can only include a linear term in $\Phi$ but since there are two $\SU(2)$ gauge groups, we find that both for 
$z=-3$ and $z=-4$ we should add three operators to the above counting. 
For $N=3$ and $N=4$ the heavy operator is relevant only for $z=-2$ and it has R-charge $\approx 1.002$ and $\approx 1.53$ respectively. 
As we have already explained, in this case our theory is $\doublewidetilde{D}_N$ and the heavy operator is obtained by setting to one 
all the $k_i$ parameters of section \ref{sec:heavy}. For $N=3$ we obtain a relevant operator as long as $\sum_i k_i<7$ whereas for $N=4$ we have the constraint $\sum_i k_i<5$. We conclude that we should add to the previous counting four relevant operators for $N=3$ and two 
relevant operators for $N=4$. In all other cases the heavy operators have R-charge larger than two.

The results are neatly summarized in the following table.
\begin{table}[!htb]
\centering
{\renewcommand{\arraystretch}{1.2}%
\begin{tabular}{@{}l  c  c  c  c c  c@{}}
& $N=2$ & $N=3$ & $N=4$ & $N=5$ & $N=6,7,8$ & $N>8$ \\
\toprule
$-z = 2$  & \multicolumn{1}{!{\vrule width 1pt} c}{ free theory} & 13 & 14 & \multicolumn{2}{|c|}{$3N$} & 24 \\

$-z = 3$ & \multicolumn{1}{!{\vrule width 1pt}c}{8} & \multicolumn{3}{|c|}{$5N-1$} & \multicolumn{2}{c}{29}\\

$-z = 4$ & \multicolumn{1}{!{\vrule width 1pt}c}{10} & 20 & \multicolumn{1}{c|}{27} & \multicolumn{3}{c}{28}\\

$4 < -z < z_\mathrm{c}(N)$ & \multicolumn{3}{!{\vrule width 1pt}c|}{$(1-N)(1+2z)$} & \multicolumn{3}{c}{$-4 -8z$} \\

$ -z \geq z_\mathrm{c}(N)$ & \multicolumn{3}{!{\vrule width 1pt} c|}{$(1-N)(1+2z)$} & \multicolumn{3}{c}{$-3 -6z$} \\
\bottomrule
\end{tabular}}
\caption{Total number of relevant operators (including the heavy ones) in the $g=0$ case.}
\label{tab:rel-sphere}
\end{table}

\section{Conclusions} 
\label{sec:conc}

In this paper we have proposed a field theoretic construction for the $\mathcal{N}=1$ inaccessible theories studied holographically in \cite{bbbw}. In that paper, the $a$ and $c$ central charges for models with negative $p$ or $q$, or genus $g\leq 1$, was correctly computed but the authors could not derive the same results from a field theory perspective. Here we have constructed the field theories corresponding to such choice of integer labels, and computed their central charges exactly (see section \ref{sub:charges-inac}). Our results correctly reproduce all formulae in \cite{bbbw}. The new field theories are of class $\mathcal{S}$, and are constructed by gluing together in an $\mathcal{N}=2$ or $\mathcal{N}=1$ way new building blocks besides $T_N$. The former are all obtained starting from a $T_N$ block and applying the nilpotent Higgsing procedure repeatedly: We first couple a new adjoint chiral $M$ to the moment map $\mu$ of $T_N$ associated to one of its punctures, and then give a maximal nilpotent vev to $M$. This completely ``closes'' the chosen puncture. This way, we can construct new building blocks with $\SU(N) \times \SU(N)$ ($D_N$), $\SU(N)$ ($\widetilde{D}_N$) or no flavor symmetry at all ($\doublewidetilde{D}_N$).

We have also derived chiral ring relations for the new blocks, starting from the known ones for $T_N$. Our results are neatly summarized in section \ref{sub:chiral-results}. Once we take these relations into account (as well as some dynamical properties of our models) we can give a simple explanation to some puzzles that were left open in the holographic analysis of \cite{bbbw} (namely, we can identify the correct ``heavy operator'' that matches the energy of an M2-brane wrapped on the Riemann surface; see section \ref{sec:heavy}). Along the way we have derived several new $\mathcal{N}=1$ dualities involving $T_N$ as well as $D_N$, $\widetilde{D}_N$, and $\doublewidetilde{D}_N$ from the original ``swap duality'' of \cite{gadde-maruyoshi-tachikawa-yan}. A dependency tree among the various dualities is depicted in figure \ref{fig:map}.

A count of relevant operators for every inaccessible model has been provided in section \ref{sec:index}, where we have also compared our results with those predicted by the superconformal index of \cite{beem-gadde}.

Finally, notice that our purely field-theoretical construction is closely related to the concept of ``discrete charges'' discussed by \cite{gaiotto-razamat} in the context of class $\mathcal{S}_k$ theories. There it was noticed that, since six-dimensional $(1,0)$ theories generically exhibit some global symmetry, when we compactify them to four dimensions we expect to be able to turn on fluxes for the global symmetry bundles, accounting for the aforementioned charges. Our models are obtained by gluing together $T_N$ and some new building blocks (see e.g. figure \ref{fig:g36Dn}). In particular, starting from the $\mathcal{N}=2$ trinion we can change the first Chern class of the trivial bundle to obtain the new blocks. This is the geometric interpretation of the nilpotent Higgsing procedure, and is roughly analogous to the introduction of discrete charges (i.e. turning on fluxes $c_1(\mathcal{F})$ for global symmetry bundles $\mathcal{F}$). For this reason we believe our method is the simplest example of a rather general construction applicable to most $(1,0)$ models, and is certainly a good starting point to get a better handle on more general cases. \newline

We will now list a few possible avenues of future investigation. In this paper we have exploited four-dimensional compactifications of the six-dimensional $(2,0)$ theory of type $A_{N-1}$, but our analysis can be extended to models obtained by compactifying that of type $D_N$. (The latter is the low-energy theory of $2N$ M5-branes probing the spacetime singularity $(\rr \times \cc^2)/\zz_2$ in eleven dimensions, and should not be confused with the four-dimensional $D_N$ block.) The basic input we need is the chiral ring of the $T_N$ theory with $\SO(2N)_A \times \SO(2N)_B \times \SO(2N)_C$ flavor symmetry (see e.g. \cite{tachikawa-6dDn,nishinaka}). Another interesting direction is the analysis of more general compactifications, in particular the case in which the total space of the Calabi--Yau three-fold is not the direct sum $\mathcal{L}_1 \oplus \mathcal{L}_2 \rightarrow \mathcal{C}_g$ of two line bundles over the Riemann surface $\mathcal{C}_g$.

Another generalization involves applying the nilpotent Higgsing procedure to the class $\mathcal{S}_k$ trinion $T_N^k$. The latter is a nonlagrangian $\mathcal{N}=1$ theory with $\SU(N)^k_A \times \SU(N)^k_B \times \SU(N)^k_C$ global symmetry, and is used as building block in the construction of $\mathcal{N}=1$ SCFT's of so-called class $\mathcal{S}_k$. This new class of theories was recently introduced in \cite{gaiotto-razamat}, and is roughly obtained by taking a $\mathbb{Z}_k$ orbifold of class $\mathcal{S}$. (The six-dimensional origin requires having a stack of $N$ M5-branes -- wrapping the Riemann surface $\mathcal{C}_g$ -- probe a spacetime singularity $\cc^2/\mathbb{Z}_k$, so that supersymmetry is reduced to $(1,0)$.) It could be possible to flip and close a puncture of $T_N^k$ as we have done for $T_N$, and obtain a new block $D_N^k$. Repeating this game for every puncture should lead us to a collection of new blocks with which we can construct new SCFT's of class $\mathcal{S}_k$, by gluing them together in an $\mathcal{N}=1$-preserving way.






\section*{Acknowledgments}
We are indebted to N.~Bobev for collaboration at an early stage of this project, and to A.~Amariti, K.~Maruyoshi and S.~Razamat for interesting discussions. M.F.~is a Research Fellow of the Belgian F.R.S.-FNRS. The work of M.F.~and S.G.~was partially supported by the ERC Advanced Grant ``SyDuGraM'', by IISN-Belgium (conventions 4.4514.08 and 4.4503.15), by the Belgian F.R.S.-FNRS (convention FRFC PDR T.1025.14), and by the ``Communaut\'e Fran\c{c}aise de Belgique" through the ARC program. M.F.~gratefully acknowledges support from the Simons Center for Geometry and Physics, Stony Brook University at which some of the work for this paper was performed during the IX Simons Summer Workshop.

\appendix

\section{New \texorpdfstring{$\mathcal{N}=1$ dualities with $T_N$}{New N=1 dualities with TN}} 
\label{app:dualities}

\begin{figure}[ht]
\centering
  \begin{minipage}{1\textwidth}
     \captionsetup{type=figure}
        \centering
        \subcaptionbox{Two $T_N$ blocks (of different sign) connected via an $\mathcal{N}=1$ vector multiplet.
        \label{fig:swap-unflipped}}[.45\columnwidth]{
        \vspace*{1.085cm}
        \includegraphics{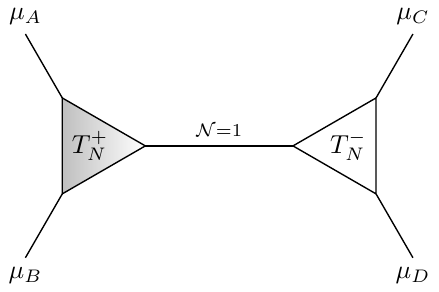}}\hspace{.5cm}
		\subcaptionbox{Two (swapped) $T_N$ blocks connected via an $\mathcal{N}=1$ vector multiplet and with flipped punctures.
        \label{fig:swap-flipped}}[.45\columnwidth]{
        \includegraphics{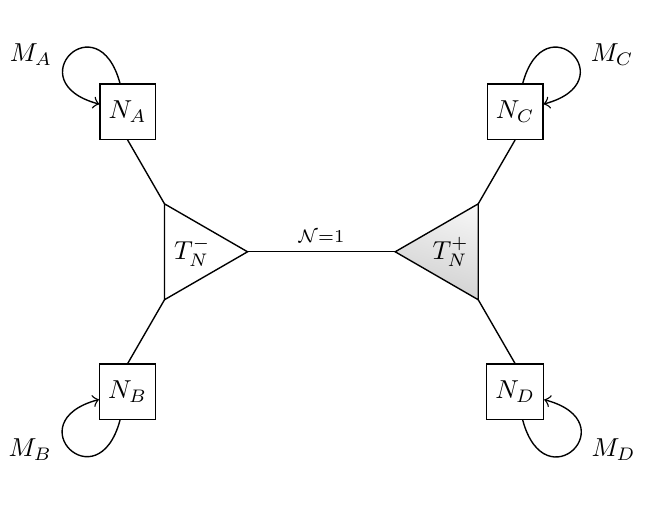}}
\end{minipage}
\caption{The original ``swap'' duality of \protect\cite[Fig. 3]{gadde-maruyoshi-tachikawa-yan} performed at the ``node'' represented by the $\mathcal{N}=1$ tube. The superpotential $\mathcal{W}= \Tr \mu_1 \mu_2$ of the theory in \ref{fig:swap-unflipped} becomes $\mathcal{W}_\mathrm{dual}= \Tr \mu_1 \mu_2 + \Tr M_A \mu_A + \Tr M_B \mu_B + \Tr M_C \mu_C + \Tr M_D \mu_D$, of the theory in \ref{fig:swap-flipped}. $\mu_{1}$ and $\mu_{2}$ are the moment maps associated with the flavor symmetries of $T_{N}^+$, respectively $T_N^-$, that are gauged together.}
\label{fig:swap}
\end{figure}

We shall now prove the dualities used in section \ref{sec:chiral} to derive new chiral ring relations for $D_N$, $\widetilde{D}_N$, and $\doublewidetilde{D}_N$. Our starting point is the so-called ``swap'' duality derived in \cite{gadde-maruyoshi-tachikawa-yan}, which is shown in figure \ref{fig:swap}.\newline

First, we prove a preliminary result. As is well known, if we couple two $T_N$ theories through an $\mathcal{N}=2$ vector multiplet and then completely close one of the punctures in either theory, we are left with a single $T_N$. If we instead couple two trinions through an $\mathcal{N}=1$ vector multiplet (and add the corresponding superpotential term $\Tr \mu_1 \mu_2$), upon closing one puncture we get a single $T_N$ block with a flipped puncture. The latter can be seen as a remnant of the Higgsed trinion. 

In order to prove this, it is convenient to first partially close a puncture to the minimal one; this turns (the partially Higgsed) $T_N$ into a hypermultiplet in the bifundamental of $\SU(N) \times \SU(N)$. We then completely close the minimal puncture. 

After the first step we get an $\SU(N)$ gauge theory coupled to $T_N$ and to $N$ hypermultiplets in the (anti-)fundamental, $Q$ and $\tilde{Q}$. The superpotential becomes:
\begin{equation}\label{eq:sup1-swap}
\mathcal{W}_{\mathcal{N}=2} = \Tr \Phi (\mu-(Q \tilde{Q})_0 )\ , \qquad \mathcal{W}_{\mathcal{N}=1} = \Tr \mu(Q \tilde{Q})_0\ ,
\end{equation}
where $(Q \tilde{Q})_0 = Q \tilde{Q} - \frac{1}{N}  Q \tilde{Q}$ denotes the traceless part. To completely close the minimal puncture we should now give a vev to the baryon built out of $Q$ (or $\tilde{Q}$), i.e. $\left\langle Q \right\rangle = 1_N$. This completely Higgses the $\SU(N)$ gauge group, whereas the $\SU(N)$ diagonal combination of flavor and gauge groups is unbroken. The $Q$ multiplet is eaten by the Higgs mechanism, and the superpotential \eqref{eq:sup1-swap} becomes:
\begin{equation}\label{eq:sup2-swap}
\mathcal{W}_{\mathcal{N}=2} = \Tr \Phi (\mu- \tilde{Q}_0 )\ , \qquad \mathcal{W}_{\mathcal{N}=1} = \Tr \mu\,\tilde{Q}_0\ .
\end{equation}
We see that in the $\mathcal{N}=2$ case both $\Phi$ and $\tilde{Q}_0$ become massive, hence the superpotential disappears. We are thus left with a single copy of $T_N$. On the contrary, in the $\mathcal{N}=1$ case we get a trinion coupled to $\tilde{Q}_0$ (which is now in the adjoint of $\SU(N)$) through the superpotential in \eqref{eq:sup2-swap}. This is nothing but $T_N$ with a flipped puncture (the ``flipping field'' being $\tilde{Q}_0$).\newline

\begin{figure}[ht]
\centering
  \begin{minipage}{1\textwidth}
     \captionsetup{type=figure}
        \centering
        \subcaptionbox{$T_N$ (of either sign) and its three moment maps.
        \label{fig:Tn-a}}[.45\columnwidth]{
        \vspace*{.6cm}
        \includegraphics{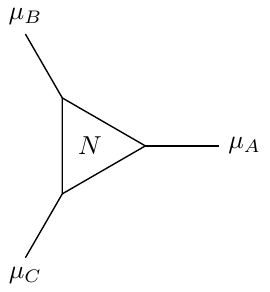}}\hspace{.5cm}
        \subcaptionbox{Another duality frame for $T_N$.
        \label{fig:Tn-b}}[.45\linewidth]{
        \includegraphics[scale=.25]{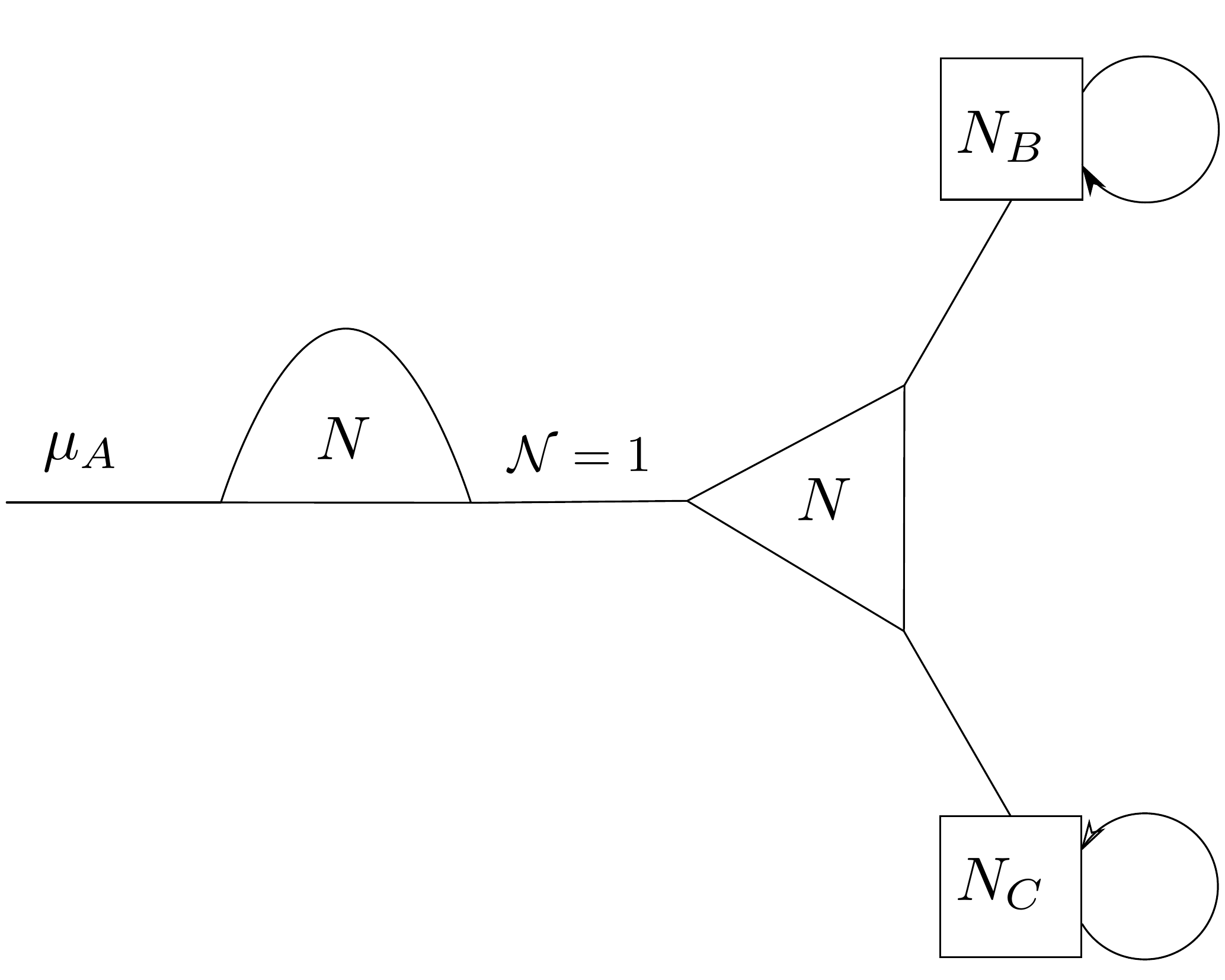}
        }
\end{minipage}
\caption{This duality is obtained starting from \protect\ref{fig:swap}, closing one puncture $X$ (either of $T_N^+$ or $T_N^-$), and then unflipping once. Following both duality frames along this process we reach \protect\ref{fig:Tn-a} and \protect\ref{fig:Tn-b}. In the latter, the two blocks have opposite signs, as required by the $\mathcal{N}=1$ gauging (the exact sign assignment depends on the choice of $X$ in \protect\ref{fig:swap}).}
\label{fig:Tn}
\end{figure}

\begin{figure}[ht]
\centering
\includegraphics[scale=.5]{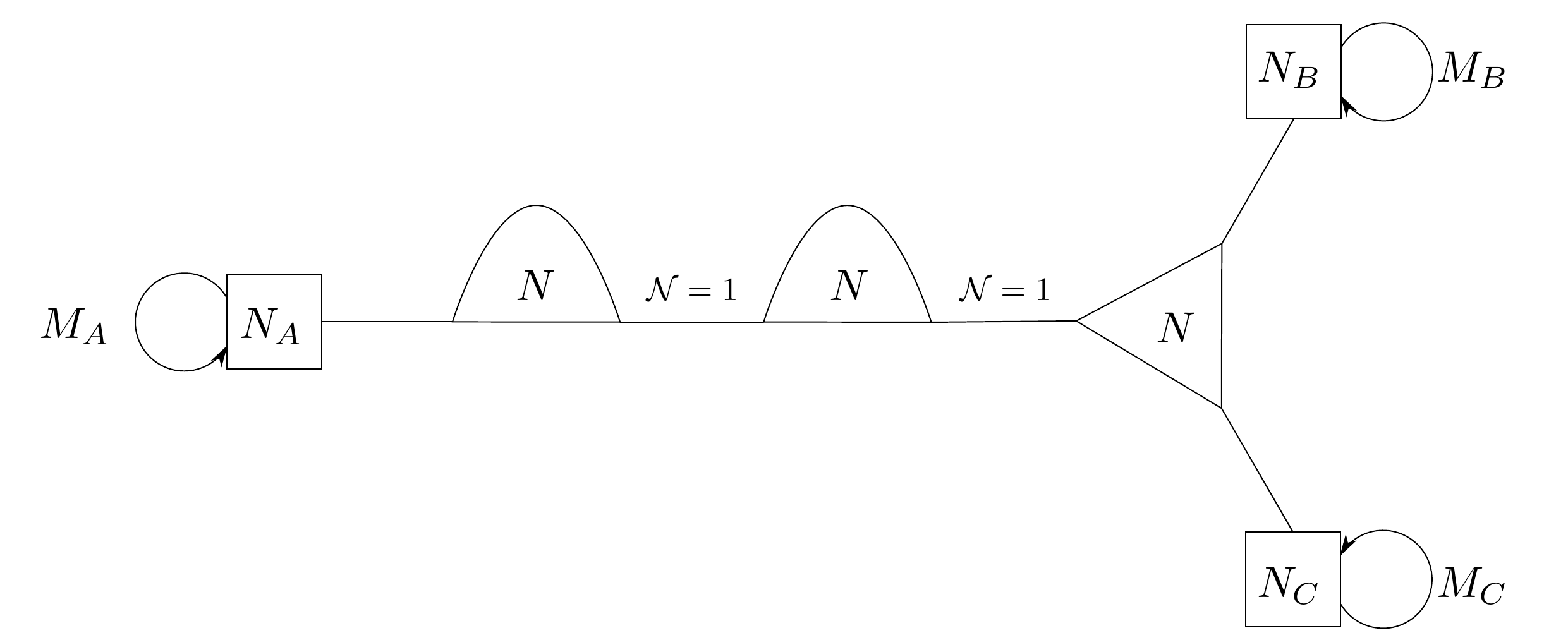}
\caption{An intermediate duality frame leading to \protect\ref{fig:Dnttdual-b} upon closing all punctures. Before closing the punctures, the duality with \protect\ref{fig:Tn3M-a} can be understood in a more direct way by noticing that two coupled $D_N$ blocks (of opposite sign) connected to $T_N$ via an $\mathcal{N}=1$ gauging is equivalent to the empty theory. (For instance, it is easy to see that the $a$ and $c$ central charges vanish due to the various opposite contributions.) It now suffices to ``visually erase'' $D_N\ \overset{\mathcal{N}=1}{\text{\textemdash}\!\text{\textemdash}} \ D_N\ \overset{\mathcal{N}=1}{\text{\textemdash}\!\text{\textemdash}}$ from the above figure to obtain \protect\ref{fig:Tn3M-a}.}
\label{fig:DnDnTn}
\end{figure}

Equipped with this result, let us consider the swap duality of \cite{gadde-maruyoshi-tachikawa-yan}, depicted in figure \ref{fig:swap}: Two $T_N$ blocks coupled through an $\mathcal{N}=1$ vector multiplet is equivalent to the same theory with four flipped punctures.
If we completely close one puncture on both sides of the duality,\footnote{The choice of puncture $X$ is arbitrary but must be consistent across the two duality frames. In figure \ref{fig:swap-unflipped} the chosen puncture is closed (completely) by giving the associated moment map $\mu_X$ a nilpotent vev (i.e. $\left\langle \mu_X \right\rangle$ is a maximal Jordan block). In figure \ref{fig:swap-flipped} we instead give $M_X$ the nilpotent vev \eqref{eq:vev}.} the theory in figure \ref{fig:swap-unflipped} reduces to $T_N$ with a flipped puncture (according to the above argument), while that in figure \ref{fig:swap-flipped} reduces to $D_N$ coupled to $T_N$ (through an $\mathcal{N}=1$ gauging) with all three punctures flipped.\footnote{Notice that $D_N$ inherits the sign of the $T_N^\pm$ theory it comes from. This is relevant whenever the former is connected via an $\mathcal{N}=1$ gauging to another block, which must then have opposite sign. By the same token, we will also have $\widetilde{D}_N$ and $\doublewidetilde{D}_N$ blocks of given sign.} Schematically:
\begin{equation}\label{eq:Tnflippedonce}
M_A\ \text{\textemdash}\ T_N\ \text{\textemdash}\ \mu_{B,C}\quad \xleftrightarrow{\ \text{dual}\ } \quad M_A\  \text{\textemdash}\ D_N\ \overset{\mathcal{N}=1}{\text{\textemdash}\unskip \text{\textemdash}}\ T_N\ \text{\textemdash}\ M_{B,C}\ .
\end{equation}
(By further closing a flipped puncture on both sides we directly obtain the duality depicted in figure \ref{fig:Dndualframes}. This proves the claim made in the caption to that figure.) We can easily obtain variants of this result by flipping (or undoing the flipping of) corresponding punctures in both duality frames. To flip we just introduce an extra chiral $M$ coupled to the moment map associated with the puncture; to unflip we introduce a second chiral $N$ coupled quadratically to the first (i.e. via $\Tr M N$). This makes both multiplets massive, effectively undoing the flipping (this is analogous to Seiberg dualizing twice). Using this procedure we immediately find that $T_N$ with all punctures flipped is equivalent to the theory in figure \ref{fig:Tn3M-b}, whereas $T_N$ is equivalent to that in figure \ref{fig:Tn-b}.
If we now replace the $T_N$ block on the right of figure \ref{fig:Tn3M-b} (with all punctures unflipped) with figure \ref{fig:Tn-b}, we find figure \ref{fig:DnDnTn}. Upon closing all punctures we get the duality depicted in figure \ref{fig:Dnttdualframes}.

\begin{figure}[!ht]
\centering
\hspace*{-.775cm}
\includegraphics[scale=1.22]{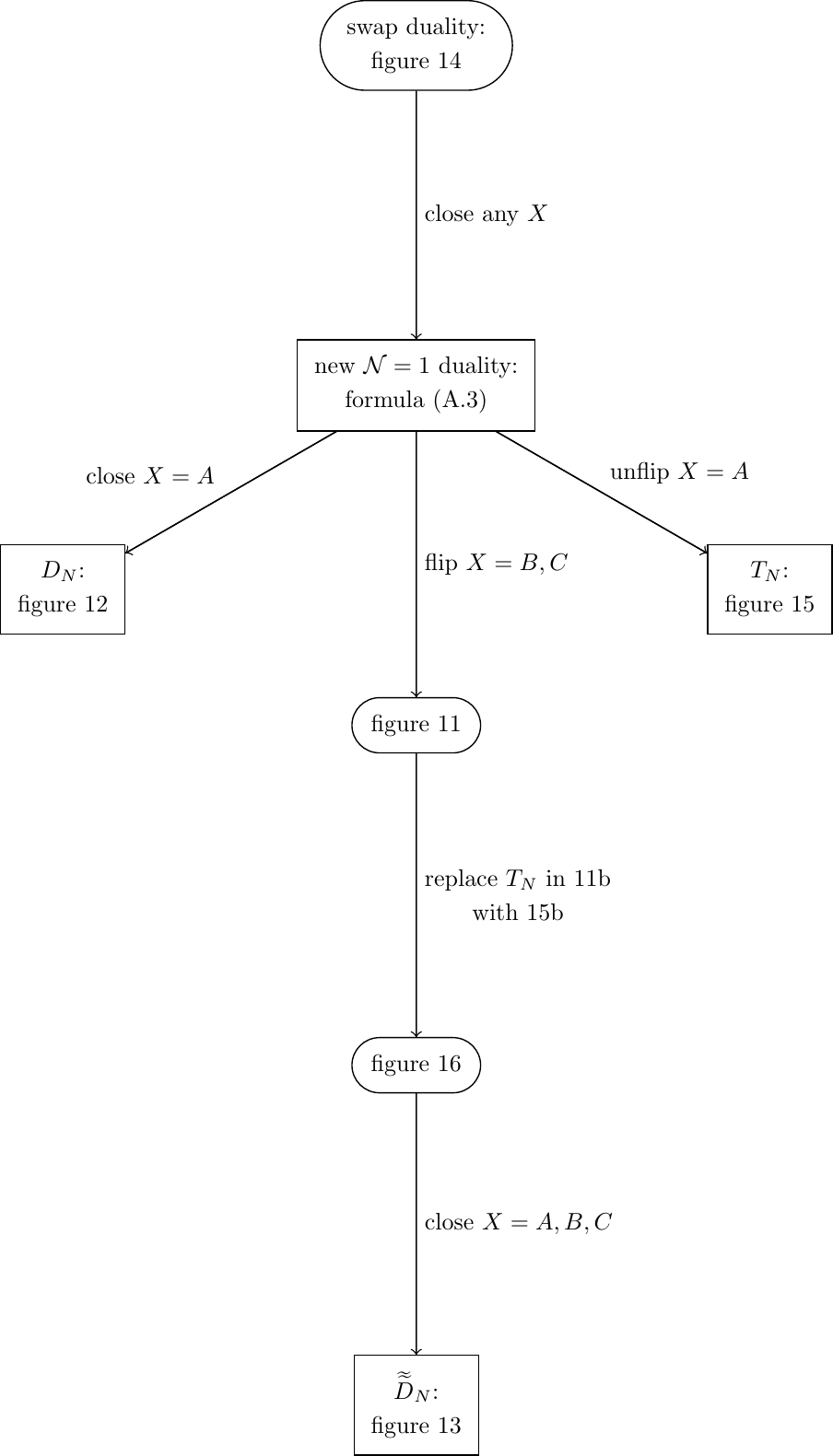}
\caption{All the dualities used in the paper descend from the swap of \protect\cite{gadde-maruyoshi-tachikawa-yan} by subsequently applying the specified moves: completely closing or (un)flipping a puncture $X$ of $T_N$ (i.e. maximally Higgsing its $\SU(N)_X$ flavor symmetry or coupling the latter to $M_X$ via $\Tr M_X \mu_X$, respectively). The proposed dualities can be checked by matching global symmetries, $a$ and $c$ central charges, and the structure of the superconformal index on both sides.}
\label{fig:map}
\end{figure}

The dependency tree depicted in \ref{fig:map} displays all the exploited dualities and summarizes the above discussion. 

\clearpage



\bibliography{neg}
\bibliographystyle{at}

\end{document}